\documentclass[graybox, nosecnum, natbib]{svmult}

\usepackage{amsfonts}
\usepackage{mathptmx}       % selects Times Roman as basic font
\usepackage{helvet}         % selects Helvetica as sans-serif font
\usepackage{courier}        % selects Courier as typewriter font
\usepackage{type1cm}        % activate if the above 3 fonts are
\usepackage{makeidx}         % allows index generation
\usepackage{graphicx}        % standard LaTeX graphics tool
\usepackage{multicol}        % used for the two-column index
\usepackage[bottom]{footmisc}% places footnotes at page bottom
\usepackage{hyperref}        %for hyperlinks
\usepackage{soul}            % for high-lighting of text
\usepackage{braket}
\usepackage{amsmath}
\usepackage{subeqnarray}
\usepackage{fancyhdr}
\usepackage{mathtools,MnSymbol}
\usepackage{bbold}
\usepackage{braket}

\makeindex             % used for the subject index
\begin{document}
\title*{Symmetry restoration methods}
\author{
Jiangming Yao,  %~\thanks{Corresponding author} 
%Peter Ring
}
%~\thanks{Corresponding author}
\institute{
J. M. Yao \at  School of Physics and Astronomy, Sun Yat-sen University,  Zhuhai, 519082 China,   \\ \email{yaojm8@sysu.edu.cn} 
}

\maketitle

\abstract{Symmetry techniques based on group theory play a prominent role in the analysis of nuclear phenomena, and in particular in the understanding of observed regular patterns in nuclear spectra  and selection rules  for electromagnetic transitions. A variety of symmetry-based nuclear models have been developed in nuclear physics, providing efficient tools of choice to interpret nuclear spectroscopic data. This chapter provides a pedagogical introduction to the basic idea of symmetry-breaking mechanism and symmetry-restoration methods in modeling atomic nuclei. }
 
% \renewcommand{\headrulewidth}{0pt}
% \thispagestyle{fancy}
% \fancyhf{}
% \rhead{}
% \lhead{}

%\tableofcontents{}

  %%%%%%%%%%%%%%%%%%%%%%%%%%%%%%%%%%%%%%%%%%%%%%%%%%%%%%%%%%%%%%
 \section{Introduction}
   Symmetries and conservation laws are indispensable keys to understanding the structure of atomic nuclei the dynamics of which are described fundamentally in terms of quarks and gluons interacting with the strong interaction governed by the theory of quantum chromodynamics. Due to the non-perturbative nature of the strong interaction in low-energy region, atomic nuclei are usually modeled in terms of nucleon degrees of freedom instead. Therefore, the symmetries of atomic nuclei are determined by the Hamiltonian composed of nucleon-nucleon interactions. Like those of many macroscopic systems, nuclear Hamiltonian possesses the symmetries  associated with geometric transformations in space-time coordinates, including translational and rotational symmetries, and space-inversion invariance. Furthermore, atomic nucleus is a quantum many-body system that is also characterized by the symmetries defined in abstract spaces, such as the spin and isospin symmetries as proposed by Wigner~\cite{Wigner:1937} and the rotation invariance in Fock space corresponding to particle-number conservation.  The exploration of these symmetries and their mathematical representations is essential to modeling atomic nuclei as it provides  not only constraints on the nucleon-nucleon interactions, but  also  guidelines on the choices of model spaces for nuclear wave functions. The latter will be the main focus of this Chapter.
    
   The exact solution to a nuclear many-body problem is computational challenging.  One of the important approaches to atomic nuclei is the interacting shell model which represents nuclear wave function as a superposition of all possible symmetry-conserving configurations.  The shell model has achieved great success in understanding the structure of atomic nuclei, but its applicability has been strongly limited by the exponential growth of the number of configurations with nucleon numbers. To mitigate the computational challenge,  symmetry techniques can be exploited to optimize model space for atomic nuclei with strong collective correlations~\cite{Launey:2016PPNP}. Alternatively,  symmetry has also been exploited to describe atomic nuclei in terms of collective degrees of freedoms, such as the collective models~\cite{Bohr:1976RMP,Mottelson:1976RMP} developed by Bohr and Mottelson, and the interacting boson model~\cite{Arima:1975PRL} developed by Arima and Iachello. In the former, the nuclear Hamiltonian is constructed in terms of collective coordinates characterizing certain shapes of the nuclear surface preserving the corresponding geometrical symmetries. In the latter,  the Hamiltonian is constructed based on the algebra of symmetry group in terms of interacting bosons of different ranks which are composite operators for pairs of correlated nucleons.   These two methods have turned out to be very successful in describing a wide variety of nuclear properties, especially nuclear spectroscopy of vibration and rotation excitations.

  The content of  this Chapter will be mainly about another type of nuclear models starting from the reference state of a mean-field calculation. Mean-field approximation has been frequently employed in modeling quantum many-body systems. An atomic nucleus is no exception. In this approximation, the atomic nucleus is described as a bound system of independent nucleons trapped in a mean-field potential determined self-consistently by themselves. The nuclear wave function becomes a single Slater determinant of single-particle or quasiparticle basis functions~\cite{Meng:2006PPNP}, and the many-body problem is reduced to an effective one-body problem that can be easily solved. It turns out that this simplified nuclear wave function is often incapable of simultaneously describing  essential correlations of an atomic nucleus and also satisfying the conservation laws. Therefore,  certain conservation law is allowed to be violated in the mean-field potential to incorporate more correlations. In this case, the state of the system  is referred to as a state of broken symmetry, and it is described by the wave function that does not possess corresponding quantum numbers. The use of such a symmetry-breaking wave function for atomic nucleus suffers from some drawbacks as the interpretation of nuclear spectroscopic data becomes obscure.   To solve this problem, symmetry-restoration methods are introduced to recover  missing quantum numbers.  This strategy is the core idea of the symmetry-projected generator coordinate method (PGCM), in which the nuclear wave function is constructed as a linear combination of non-orthogonal basis functions that are projected out from ``deformed" mean-field wave functions~\cite{Bender:2003RMP}. Following this idea, many advanced nuclear models have been established, including the multi-reference energy-density-functional (MR-EDF) method~\cite{Niksic:2011PPNP,Egido:2016PS,Robledo:2018JPG}, Monte-Carlo shell model (MCSM)~\cite{Otsuka:2001PPNP} and the projected shell model (PSM) based on either a schematic pairing-plus-qudrupole Hamiltonian~\cite{Hara:1995}, or the effective interaction derived from a covariant density functional theory~\cite{Zhao:2016PRC,Wang:2022PRC}.  
  These methods have been frequently employed to interpret nuclear spectroscopic data in different mass regions.   A detailed introduction to symmetry breaking in nuclear mean fields and the restoration of broken symmetries with projection techniques can be found in  textbooks~\cite{Ring:1980,Blaizot:1986}, and in the recent review papers~\cite{Sheikh2021_JPG48-123001,Yao:PPNP} and the references therein.  Recently, this idea has also been implemented into {\em ab initio}  studies of atomic nuclei based on the {\em many-body expansion methods}, including the symmetry broken and restored coupled-cluster theory~\cite{Duguet:2015EJPA-1,Duguet:2015EJPA-2,Signoracci:2015PRC,Hagen:2022PCC}, in-medium generator coordinate method (IM-GCM)~\cite{Yao:2018wq,Yao:2020PRL} -- a new variant of multi-reference in-medium similarity renormalization group method~\cite{Hergert:2016PR}, and many-body perturbation theory (MBPT)~\cite{Frosini:2021a,Frosini:2021b,Frosini:2021c}.

This Chapter provides a pedagogical introduction to the symmetries of nuclear Hamiltonian that are allowed to be broken in nuclear mean-field potentials to incorporate many-body correlations, with special emphasis on  the methods that have been frequently employed to restore these broken symmetries to achieve an accurate description of nuclear-structure and decay properties. Some illustrative applications to selected nuclear-structure problems are also discussed. The frontier and recent exciting progress in modeling atomic nuclei based on symmetry techniques are briefly mentioned.

\subsection{Symmetry and group representations}

Let us start with the basic knowledge of symmetry and group theory. Considering a set of transformations $G=\{\hat e, \hat g_1, \hat g_2, \cdots \}$, the set $G$ forms a group if the following conditions are fulfilled for all the elements $g_i$ belonging to the group~\cite{Stancu:1991,Frank:2009}
 \begin{subequations}
 \begin{align}
 {\rm identity}: & \quad \hat e\hat g_i  =\hat g_i\hat e  = \hat g_i,  \\
 {\rm inverse}:  & \quad  \hat g_i g^{-1}_i =  \hat e, \\
 {\rm closure}:  & \quad  \hat g_i\hat g_j = \hat g \in G, \\
 {\rm associativity}: & \quad (\hat g_i \hat g_j) \hat g_k = \hat g_i(\hat g_j\hat g_k).
\end{align}  
 \end{subequations}
In the group $G$, the element $\hat e$ is the {\em unity} or {\em identity} element, and $\hat g^{-1}_i$ is the inverse of the element $\hat g_i$. The group $G$ is called a finite group  if the number of the element is finite, and this number is called the {\em order} of the finite group. There are also groups with an infinite number of elements. These groups can be either the {\em discrete} groups of infinite order or the {\em continuous} groups. For the latter, the group element denoted as $\hat R(\mathbf{\varphi})$ depends on a finite set of continously varying parameters $\mathbf{\varphi}=\{\varphi_1, \varphi_2, \cdots, \varphi_r\}$.

In quantum mechanics, group theory offers a systematic way of finding the properties of eigenstates under various transformation from the symmetries of Hamiltonian.  The eigenstates form linear spaces providing matrix representations of the group transformations $G$. Specifically, for the Hamiltonian $\hat H_0$ of a quantum system which is invariant under the transformations $\hat g$,  one has 
\begin{equation}
[\hat g, \hat  H_0]=0, \quad \forall \hat g \in G.
\end{equation} 
 Considering a linear space $L$ composed of $N$ orthonormal basis states $\{|\Phi_1\rangle$, $|\Phi_2\rangle$, $\cdots$, $|\Phi_N\rangle\}$, any group element $\hat g$ acting on the states induces a $N\times N$ matrix $D_{ji}(g)$, i.e.,
\begin{equation}
\hat g|\Phi_i\rangle =\sum^{N}_{j=1} D_{ji}(g)\left|\Phi_j\right\rangle \in L,\quad {\rm for}\quad  \forall|\Phi_i\rangle \in L, \quad {\rm and} \quad \forall \hat g \in G,
\end{equation}
where the indices $i, j$ run from 1 to $N$. The $N\times N$ matrices $D(g)$ with the element determined by $D_{ji}(g)=\bra{\Phi_j}\hat g\ket{\Phi_i}$ preserve the multiplicative structure of the group $G$ and form a matrix representation of the group. The group $G$ can have a different matrix representation denoted as $D^{\prime}(g)$ if a different linear space $L'$ is employed. If the two representations $D(g)$ and $D^{\prime}(g)$ are connected by a {\em similarity} transformation  $X$ for all the group elements, these two representations are called {\em equivalent}. By applying different similarity transformations $X^{(i)}$, one can generate many equivalent representations $D^{(i)}$, among  which, if there is one with the following form,
\begin{eqnarray}
\label{eq:representation-transform}
X^{-1}D(g)X
=\begin{bmatrix}
 D^{\lambda_1}(g) & T^{12}(g)  \\
 0 &D^{\lambda_2}(g)  \\ 
\end{bmatrix},\quad  {\rm for}\quad\forall \hat g\in G,
\end{eqnarray}
then, the representation $D(g)$ is called {\em reducible}. The block matrices $D^{\lambda_1}(g)$ and $D^{\lambda_2}(g)$ are two square matrices with dimensions $d_{\lambda_1}$ and $d_{\lambda_2}$, respectively. It indicates that there is  a subspace $L^1\subset L$,  $\forall |\Phi^{\lambda_1}_l\rangle \in L^1$ is transformed into another state $\hat g|\Phi^{\lambda_1}_l\rangle$ which also belongs to the subspace, and this is true for all the transformations of the group, namely,
\begin{eqnarray}
\hat g|\Phi^{\lambda_1}_l\rangle = \sum_{k}D^{\lambda_1}_{kl}(g) |\Phi^{\lambda_1}_k\rangle,\quad  {\rm for}\quad\forall \hat g\in G.
\end{eqnarray}
The subspace $L^1$ fulfilling the above requirement is called {\em invariant subspace}. Furthermore, if the matrix $T^{12}(g)=0$ in (\ref{eq:representation-transform}), the representation $D(g)$ is called {\em fully reducible}.
In this case, the full $L$ space can be written as a sum of two invariant subspaces $L=L^1\oplus L^2$ and the representation 
\begin{equation} 
D(g)=D^{\lambda_1}(g)\oplus D^{\lambda_2}(g). 
\end{equation}
This property has been frequently exploited to split a large Hilbert space into a set of independent subspaces with smaller dimensions. The representation $D^\lambda(g)$ is called {\em irreducible} if there is no   similarity transformation which brings the matrices $D^\lambda(g)$ into block diagonal form simultaneously for $\forall \hat g\in G$. The {\em irreducible} representation is sometimes abbreviated as {\em irrep} for brevity. It is worth mentioning that an irrep of one full group may be reducible for its subgroup. For example, the spherical harmonics $Y_{\ell m}$ with $m=-\ell, \cdots, \ell$ forms an invariant subspace of $2\ell+1$ dimension for the rotational group  SO(3)  with the irrep of $D^\ell(g)$, which is however reducible for the subgroup  SO(2)  related to the rotation $\hat R_z(\varphi)$ in two dimensional space (along $z$-axis)
\begin{equation} 
 \hat R_z(\varphi)
 =e^{-i\varphi \hat L_z},\quad \hat L_z=-i\hbar \frac{d}{d\varphi}.
\end{equation}
In other words, the irrep of $D^\ell(R_z)$ can be further reduced into a block diagonal form
\begin{eqnarray} 
D^\ell(R_z)
=\begin{bmatrix}
 e^{im\varphi} & 0 \\
 0   & e^{-im\varphi}
\end{bmatrix},\quad \forall \hat R_z(\varphi)\in {\rm SO}(2).
\end{eqnarray}
The basis functions form {\em multiplets} of the orbital angular momentum, each of which constitutes a one-dimension irreducible invariant subspace for the SO(2) group. 

 If the group $G$ is a finite group or a {\em compact Lie} group like the orthogonal group O($n$) and the unitary group U($n$), it can be proven that any matrix representation of the group $G$ is equivalent to a {\em unitary} representation~\cite{Stancu:1991} in which the matrices are unitary for all group elements. Supposing $D^{\lambda}_{ij}(g)$  and  $D^{\lambda'}_{ij}(g)$ are two unitary irreps of the group element $\hat{R}(g)$, where the matrix element is determined by
\begin{equation}
\label{eq:irreps_SO3}
    D_{k l}^{\lambda(\lambda')}(g)=\langle \Phi^{\lambda(\lambda')}_{k}|\hat g|\Phi^{\lambda(\lambda')}_{l}\rangle,
\end{equation}
with  the indices $k, l$ running from 1 to $d_\lambda$,  one has the orthogonality relation
\begin{equation}
\label{eq:Schur1}
  \frac{1}{n_{G}}\sumint dg  D_{k l}^{\lambda}(g) [D_{k' l'}^{\lambda'}(g)]^\ast=0,
\end{equation}
for non-equivalent representations and 
\begin{equation}
\label{eq:Schur2}
   \frac{1}{n_G}\sumint dg  D_{k l}^{\lambda}(g) [D_{k'l'}^{\lambda'}(g)]^\ast=\frac{\delta_{\lambda\lambda'}}{d_\lambda}X_{kk'}X^{-1}_{l'l},
\end{equation}
for equivalent representations. The symbol $\sumint dg$ stands for the summation or integral over all the elements of the group $G$, and $n_G$ is the order (the number of elements or the volume) of the group, $d_\lambda$ is the dimension of the irrep, and $X$ is the similarity transformation which connects the two irreps of $D^{\lambda}$ and $D^{\lambda'}$. If the two irreps  are identical, then the similarity transformation is just a unity matrix, i.e., $X=\mathbb{1}$.

\subsection{Symmetry breaking in mean-field approximations}
Symmetries can be classified  according to the order of the groups, including  the symmetries associated with discrete groups (such as parity, charge conjugation, time reversal, permutation symmetry) consisting of a finite number of group elements, and the symmetries associated with continuous  groups  (such as translations and rotations symmetries in space-time or gauge angle coordinate system) consisting of an infinite number (continuum) of elements. According to Noether's theorem~\cite{Noether:1971}, {\em any continuous symmetry of the action (Lagrangian or Hamiltonian) of a physical system with conservative forces has a corresponding conservation law.} The above-mentioned symmetries imply the conservation of total momentum (energy), angular momentum, and particle numbers in atomic nuclei, respectively. On the other hand,
symmetries can also be classified according to the space where the symmetries are defined, including  the symmetries defined in the space-time coordinate system and the symmetries defined in abstract Hilbert spaces.

The symmetry of a physical system could be broken explicitly or spontaneously. The former is due to the presence of an additional interaction term that is not invariant under the symmetry transformation, such as the isospin SU(2) symmetry in atomic nuclei which breaks into SO(2) symmetry. These symmetries are called {\em dynamical} symmetries in the sense that the nuclear Hamiltonian breaks the symmetries but preserves the symmetries defined by their subgroups~\cite{Frank:2009}. The breaking of this type of symmetries in atomic nuclei is not necessary to be restored as it is determined by the nature of nuclear force. In the latter case, the Hamiltonian is kept to be invariant under the symmetry transformation, but the physical state chooses one of the multiple degenerate (ground) states which are not the eigenstates of (but connected by) the generator operator of the symmetry group. This spontaneous symmetry breaking  phenomenon could be caused by the presence of external perturbations and the thermodynamic limit. In fact, any other choice of the solutions would have exactly the same energy, which for continous symmetry implies the existence of a massless {\em Nambu–Goldstone} boson. The spontaneous symmetry breaking usually happens in  {\em infinite} systems studied in condensed matter or high-energy physics, such as the superconductors, ferromagnets, and the Higgs mechanism that generates the masses of elementary particles~\cite{Brauner:2010}. In nuclear physics, the atomic nucleus is a finite system, for which, the phenomenon of spontaneous symmetry breaking arises only as a result of approximations.  That is,  if the nuclear many-body problem is solved exactly,  symmetry should not be broken in nuclear states. This is exactly the reason why one needs to restore the symmetries broken by the employed approximations in nuclear models.

 In the self-consistent mean-field approaches, such as Hartree-Fock (HF) or Hartree-Fock-Bogoliubov (HFB), a single Slater determinant is chosen as the trial wave function of an atomic nucleus which is determined with the variational principle, that is, the minimization of the expectation value of the Hamiltonian to the wave function. The variational principle which guarantees the solution corresponding to the local energy minimum within the restricted Hilbert space cannot ensure the symmetry structure of the Hamiltonian.  As a result, the symmetries of a given nuclear Hamiltonian are not fully retained in the solutions. If one constrains the solution to preserve a certain symmetry throughout the variational calculation, one may end up with a solution that does not incorporate sufficient correlations of atomic nuclei and cannot achieve a satisfying description of nuclear basic quantities, such as binding energies.  Therefore, the mean-field approaches which allow for the breaking of the symmetries of the Hamiltonian in the mean-field potentials (also referred to as {\em symmetry-breaking mechanism}) during the variational procedure have been widely employed in nuclear  physics.  One of the celebrated successes of these symmetry-breaking mean-field approaches is shown in the description of nuclear deformation and pairing correlations~\cite{Ring:1980}.

 In molecular physics, an interplay between degenerate electronic states (single-particle motions) and a molecular vibration  (collective motions)  gives rise to a spontaneous breaking of a molecular symmetry. This is the so-called {\em Jahn-Teller} effect~\cite{Jahn:1937}, which turns out to be a general feature of quantum many-body systems induced by a linear coupling between their microscopic and collective degrees of freedom~\cite{Reinhard:NPA1984}.  In open-shell nuclei,  the degeneracy of eigenvalues of a (spherical) nuclear single-particle Hamiltonian  around the Fermi level leads to the solution unstable with respect to shape vibrations. Additional correlation energies can be gained if the  degeneracy of the single-particle Hamiltonian is lifted by allowing the breaking of the symmetry. As a result, a ``deformed" solution emerges in the variational calculation. 
 
\begin{figure}[t]
\centering
\includegraphics[width=\textwidth]{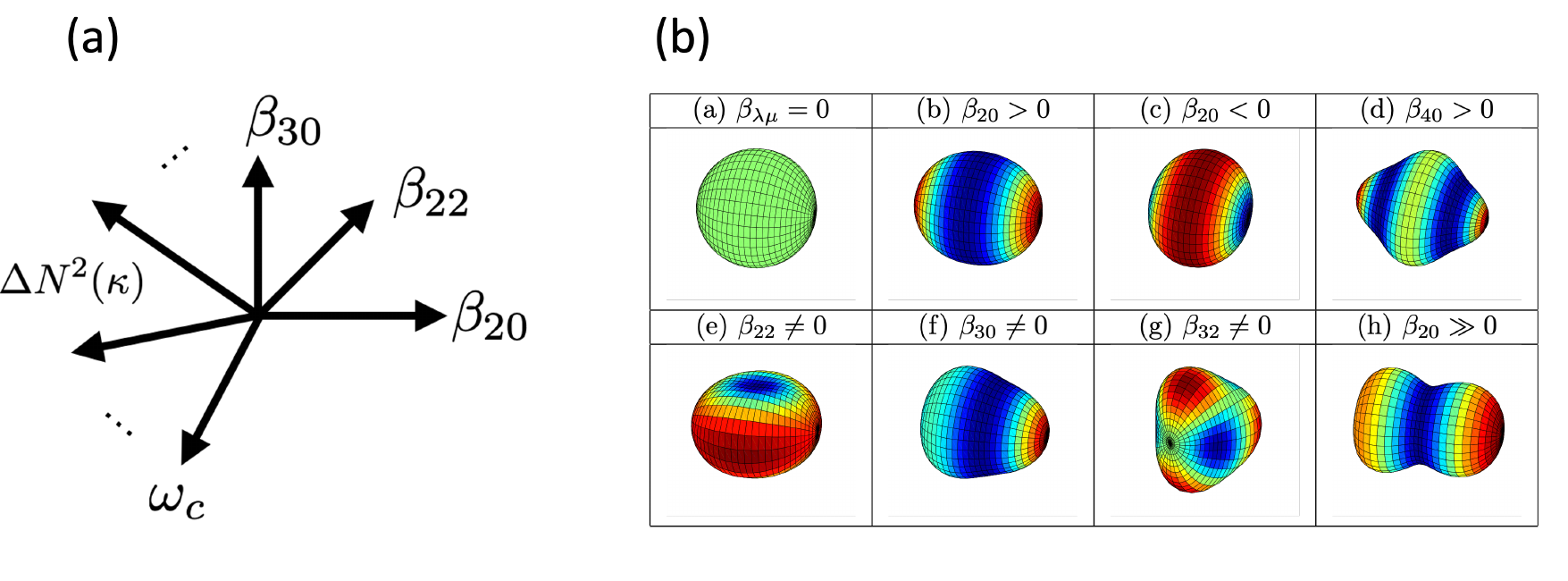}
\caption{(a) A schematic representation of a general nuclear wave function $\ket{\Psi}$ expanded in terms of non-orthogonal basis functions $\ket{\Phi(\mathbf{q})}$ in multi-dimensional collective space labeled with the coordinate vector $\mathbf{q}=\{\beta_{\lambda\mu}, \kappa, \cdots, \omega_c\}$. (b) A schematic show of some typical nuclear shapes characterized with multipole deformation parameters $\beta_{\lambda\mu}$, taken from Ref.~\cite{Lu:2012EPJ}. Fig.(b) reprinted with the courtesy of B. N. L\"{u}.}
\label{fig:generate_nuclear_states}
\end{figure}

 The obtained  wave function $\Phi(\mathbf{r}_1,\cdots,\mathbf{r}_A;\mathbf{q})$ abbreviated as $\Phi(\mathbf{q})$ of a $A$-body nuclear system from for instance the HFB approach can be generally labeled with a set of {\em collective coordinates} denoted 
 with the symbol $\mathbf{q}$, see Fig.~\ref{fig:generate_nuclear_states}(a), which includes for instance the pairing amplitude $\kappa$ or particle-number fluctuation, 
  \begin{equation}
  \Delta N^2 \equiv \bra{\Phi(\mathbf{q})} (\hat N- N_0)^2 \ket{\Phi(\mathbf{q})},\quad \bra{\Phi(\mathbf{q})}\hat N\ket{\Phi(\mathbf{q})}=N_0,
\end{equation} 
 and the dimensionless multiple deformation parameters,
 \begin{equation}
\beta_{\lambda\mu} =\frac{4 \pi}{3 A R^{\lambda}} \sum_{ij}\left(Q_{\lambda\mu}\right)_{ij} \bra{\Phi(\mathbf{q})}  a^\dagger_i a_j \ket{\Phi(\mathbf{q})},
\end{equation}
where the particle-number operator $\hat N=\sum_i c^\dagger_i c_i$ with $c^\dagger_i, c_i$ being the single-particle creation and annihilation operators in a certain basis, respectively, and the multipole moment tensor operator $Q_{\lambda\mu}=r^{\lambda} Y_{\lambda \mu}$.  The nuclear size $R=1.2 A^{1/3}$~fm, with $A$ being the mass number of the nucleus. The value of $\mu=-\lambda,-\lambda+1,\ldots,\lambda$. The onset of nonzero value of pairing amplitude $\kappa$ or $\Delta N^2$ is the result of the mixture of single-particle creation and annihilation operators, leading to the violation of gauge rotation symmetry described by the U(1) group which will be introduced in detail later.  Fig.~\ref{fig:generate_nuclear_states}(b) displays some cartoon pictures for nuclear shapes characterized by the deformation parameter $\beta_{\lambda\mu}$. 
\begin{itemize}
    \item The spherical shape corresponds to  $\beta_{\lambda\mu}=0$ for all values of $\lambda, \mu$. 
    \item The positive and negative value of $\beta_{20}$ corresponds to prolate and oblate deformed shapes, respectively. Nuclear triaxiality is defined by the nonzero value of $\beta_{22}$. The occurrence of a nonzero quadrupole deformation $\beta_{2\mu}$ induces the mixture of single-particle states with different angular momentum, leading to the violation of rotational SO(3) symmetry -- a special orthogonal group in three dimensions.
    \item If the $\beta_{3\mu}\neq0$ for $\ket{\Phi(\mathbf{q})}$, the density distribution of the solution $\ket{\Phi(\mathbf{q})}$ has a reflection-asymmetric shape. The occurrence of nonzero octupole moments induces the mixture of single-particle states with positive and negative parities. The resultant wave function  $\ket{\Phi(\mathbf{q})}$ is thus not an eigenstate of the parity operator $\hat P$.
\end{itemize}
 Therefore, the wave function $\ket{\Phi(\mathbf{q})}$ from the mean-field calculations usually mixes different irreps $\ket{\Phi_{\alpha}(\mathbf{q})}$ of symmetry groups that may include the gauge  U(1) symmetry, the rotational SO(3) symmetry,  the space-reversal $\mathbb{Z}_2$ symmetry, etc. Thus, the mean-field wave function can be generally decomposed as follows,  
\begin{equation}
\label{eq:HFB_wf}
     \ket{\Phi(\mathbf{q})} = \sum_{\alpha} c_{\alpha}(\mathbf{q}) \ket{\Phi_{\alpha}(\mathbf{q})}.
\end{equation}
where the symbol $\alpha=\{JNZ\pi,\cdots\}$ stands for a set of quantum numbers, such as angular momentum $J$, particle numbers ($N, Z$), parity $\pi$, etc. The methods to extract the component $\ket{\Phi_{\alpha}(\mathbf{q})}$ with the correct quantum numbers $\alpha$ will be introduced subsequently.

\section{Symmetry restoration methods}

Despite the great success of mean-field approaches for  nuclear ground state, the deficiency of using a symmetry-breaking wave function $\ket{\Phi(\mathbf{q})}$ in (\ref{eq:HFB_wf}) shows up in the analysis of nuclear spectroscopic properties. The restoration of missing quantum numbers in nuclear wave functions due to the admixture of different irreps is essential in the interpretation of spectroscopic data to ensure the selection rules.  This can be done by projecting out the irreps $\ket{\Phi_{\alpha}(\mathbf{q})}$ of the symmetry groups from the ``deformed" wave function in (\ref{eq:HFB_wf}). This procedure is usually called {\it symmetry restoration} in nuclear physics. 

\subsection{Generator coordinate method (GCM)}

If the Hamiltonian $\hat H_0$ is invariant under the transformation $\hat R(\mathbf{\varphi})$, or equivalently, $[\hat H_0, \hat R(\mathbf{\varphi})]=0$, then all the rotated  mean-field states are degenerate 
\begin{equation} 
      \bra{\Phi(\mathbf{q})}\hat R^\dagger(\mathbf{\varphi})\hat H_0\hat R(\mathbf{\varphi})\ket{\Phi(\mathbf{q})} = \bra{\Phi(\mathbf{q})}  \hat H_0\ket{\Phi(\mathbf{q})}.
\end{equation} 
Therefore, a natural way to restore the symmetry $G$ associated with this rotation is to use a linear superposition of a family of the rotated states for all  $\hat R(\mathbf{\varphi})\in G$,
\begin{equation} 
\label{eq:projection_trial_wf}
     \ket{\Phi_\alpha(\mathbf{q})} =  \int d\mathbf{\varphi} F^\alpha(\mathbf{\varphi})  
     \hat R(\mathbf{\varphi})\ket{\Phi(\mathbf{q})}.
\end{equation}
 This is essentially the basic idea of the generator coordinate method (GCM)~\cite{Hill:1953}, and the rotation parameter $\varphi$ can be interpreted as a {\em generator coordinate} which generates the wave functions in different orientations. The use of the trial wave function in (\ref{eq:projection_trial_wf}) with the unknown expansion coefficient $F^\alpha(\mathbf{\varphi})$ as the variational parameter is equivalent to the projection of the states with good symmetry~\cite{Blaizot:1986}. It will be shown later that the $F^\alpha(\mathbf{\varphi})$ can be derived analytically from the group theory which defines the so-called {\em projection operator}.
 
 %It will be shown later that the  $F^\alpha(g)$ is determined by different irreps of the symmetry group labeled with the symbol $\alpha$.
 
 This procedure can be carried out in two different schemes. In the {\em projection-after-variation} (PAV) scheme, both the mean-field  wave function $\ket{\Phi(\mathbf{q})}$ and the coefficient $F^\alpha(\mathbf{\varphi})$ are changing simultaneously to search for the minimum of the energy expectation to the symmetry-projected wave function $\ket{\Phi_\alpha(\mathbf{q})}$.  In contrast,  the wave function  $\ket{\Phi_\alpha(\mathbf{q})}$ in the {\em variation-after-projection} (VAP) scheme is extracted from the mean-field wave function $\ket{\Phi(\mathbf{q})}$ determined after the variational calculation. In other words, the variation of $\ket{\Phi(\mathbf{q})}$ and  $F^\alpha(\mathbf{\varphi})$ is carried out sequentially. The parameter space in the VAP is generally larger than that in the PAV, thus the VAP is expect to provide a closer-to-exact solution.  In particular, the VAP scheme can prevent the occurrence of {\em sharp} phase transition in the mean-field solution of atomic nuclei due to the artificial symmetry breaking~\cite{Mang:1976,Ring:1980}. However, the VAP scheme is more computation demanding and is thus frequently employed for parity and particle-number projections, scarcely for angular momentum projection. 
 
 In many applications,  the wave function by either the PAV or VAP scheme provides a good starting point of a more advanced nuclear model. According to GCM,  nuclear  wave function is constructed as a linear superposition of all the symmetry-projected wave functions $\ket{\Phi_\alpha(\mathbf{q})}$,
 \begin{equation}
\label{eq:GCM_wf}
     \ket{\Psi_\alpha} = \int d\mathbf{q} f_{\alpha}(\mathbf{q}) \ket{\Phi_{\alpha}(\mathbf{q})}.
\end{equation}
Here the collective coordinate $\mathbf{q}$ is the {\em generator coordinate} which is integrated out and does not appear in the final GCM wave function. The coefficient $f_{\alpha}(\mathbf{q})$, also known as collective wave function or generator function, is determined by the variational principle which leads to the following integral equation, referred to as Hill-Wheeler-Griffin equation~\cite{Hill:1953,Griffin:1957},
 \begin{equation}
\label{eq:HWG}
     \int d\mathbf{q}' \Bigg[\bra{\Phi_{\alpha}(\mathbf{q})} \hat H \ket{\Phi_{\alpha}(\mathbf{q}')} 
     - E_\alpha \bra{\Phi_{\alpha}(\mathbf{q})}  \Phi_{\alpha}(\mathbf{q}')\rangle\Bigg] f_{\alpha}(\mathbf{q}')=0.
\end{equation} 
It is worth pointing out that  the projected wave functions $\ket{\Phi_{\alpha}(\mathbf{q})}$ do not form an orthogonal basis and thus the weights  $f_{\alpha}$ are not orthogonal functions. In most cases, the function $g_{\alpha}(\mathbf{q})$ is introduced as follows,
 \begin{equation}
 \label{eq:wf_g}
g_{\alpha}(\mathbf{q})
=\int d\mathbf{q}'  \left[ \bra{\Phi_\alpha(\mathbf{q})}\Phi_\alpha(\mathbf{q}')\rangle\right]^{1/2} f_{\alpha}(\mathbf{q}'),\quad \int d\mathbf{q}|g_{\alpha}(\mathbf{q})|^2=1,
 \end{equation} 
 which fulfills the normalization condition, and reflects how configurations mix in the GCM wave function $\ket{\Psi_\alpha}$.  
 
 To distinguish the procedure of only performing a symmetry-restoration calculation with projection operators (even though this procedure also belongs to GCM) from the procedure of also mixing the projected wave functions with different collective coordinate $\mathbf{q}$, the former procedure is referred to as {\em symmetry restoration or projection}, while the latter procedure is called symmetry-projected GCM (PGCM).

  \subsection{Construction of projection operators}
   
   \subsubsection{Basic properties of a projection operator }
A straightforward way to extract the component $\ket{\Phi_\alpha(\mathbf{q})}$ with the quantum numbers $\alpha$ is by means of projection operators.  Supposing $\ket{\Phi_\alpha(\mathbf{q})}$ provides a basis function of unitary irrep of the group $G$, a projection operator $\hat P^\lambda$ is defined in such a way that 
  \begin{equation}
     \hat P^\lambda  \ket{\Phi_\alpha(\mathbf{q})}=\delta_{\lambda\alpha}  \ket{\Phi_\lambda(\mathbf{q})}.
 \end{equation}
 A repeated use of the projection operator would not change the result,
   \begin{equation}
     (\hat P^\lambda)^2  \ket{\Phi_\alpha(\mathbf{q})}
     =\delta_{\lambda\alpha}  \hat P^\lambda   \ket{\Phi_\lambda(\mathbf{q})} =\delta_{\lambda\alpha}  \ket{\Phi_\lambda(\mathbf{q})}.
 \end{equation}
  It gives rise to the important properties of the projection operator (idempotent and Hermitian) 
    \begin{equation}
    (\hat P^\lambda)^2=\hat P^\lambda, \quad \hat P^{\lambda\dagger}=\hat P^\lambda.
 \end{equation}
 Because the basis functions belonging to different representation spaces are orthogonal to each other, i.e., $\bra{\Phi_\lambda}\Phi_\alpha\rangle=\delta_{\lambda\alpha}$, the projection operator $\hat P^\lambda$ can be constructed formally in terms of the basis function as below
   \begin{equation}
     \hat P^\lambda = \sum_\sigma \ket{\Phi^\sigma_\lambda}\langle\Phi^\sigma_\lambda|,
 \end{equation}
 where $\sigma$ denotes all the set of quantum numbers other than $\lambda$.

%   \subsubsection{The L\"owdin method}
%  Supposing $\hat X$ stands for the {\em infinitesimal  generator} operator of a continuous Lie group $G$, and the component $\ket{\Phi_n(\mathbf{q})}$ of the mean-field wave function $\ket{\Phi(\mathbf{q})}$ in (\ref{eq:HFB_wf}) with $n\in\alpha$ is one of its eigenfunctions with the eigenvalue $x_n$, one has the following eigenvalue equation
%  \begin{equation}
%      \hat X \ket{\Phi_n(\mathbf{q})}=x_n\ket{\Phi_n(\mathbf{q})}.
%  \end{equation}

%  If all the eigenvalues $x_k$ of the generator operator $\hat X$ are known, one can construct the projection operator $\hat P^m$ in terms of these eigenvalues as proposed by  L\"owdin ~\cite{Lowdin:1964RMP},
%   \begin{equation}
%   \label{eq:Lowdin_projection}
%      \hat P^m =  \prod_{k\neq m} \frac{\hat X -x_k}{x_m-x_k}.
%  \end{equation}
% Based on the  fact that  the product $\prod_{k\neq m} \frac{x_n -x_k}{x_m-x_k}$ is zero for $n\neq m$, one can prove immediately that the operator $\hat P^m$ in (\ref{eq:Lowdin_projection}) projects out the component $\ket{\Phi_m}$ from the wave function $\ket{\Phi(\mathbf{q})}$,
%   \begin{equation}
%      \hat P^m   \ket{\Phi} 
%   \label{eq:Lowdin_projection_comp}
%      = \sum_n c_n \prod_{k\neq m} \frac{\hat X -x_k}{x_m-x_k}\ket{\Phi_n}
%      =\sum_n c_n \prod_{k\neq m} \frac{x_n -x_k}{x_m - x_k}\ket{\Phi_n}
%      =c_m \ket{\Phi_m}.
%  \end{equation}

  \subsubsection{The projection operator from group theory}
In group theory, the projection operator $\hat{P}_{\mu\mu}^{\lambda}$  is introduced based on the  orthogonality relation (\ref{eq:Schur2}) of the irreps $D_{\mu\mu}^{\lambda}(g)$ of  the group $G$ \cite{Cornwell:1997}
\begin{equation}
\label{eq:projection-general}
\hat{P}_{\mu\mu}^{\lambda} \equiv \frac{d_{\lambda}}{n_{G}}\int d\mathbf{\varphi} D_{\mu\mu}^{\lambda\ast}(\mathbf{\varphi}) \hat{R}(\mathbf{\varphi}),
\end{equation}
where the symbol $\mu$ is introduced to distinguish basis functions of the invariant subspace $\lambda$. The projection operator (\ref{eq:projection-general}) has the following properties
\begin{equation}
\hat{P}_{\mu\mu}^{\lambda\dagger} = \hat{P}_{\mu\mu}^{\lambda},\quad  
\hat{P}_{\mu\mu}^{\lambda}\hat{P}_{\mu\mu}^{\lambda}= \hat{P}^\lambda_{\mu\mu}.
 \end{equation}
 One can prove that the projection operator $\hat{P}_{\mu\mu}^{\lambda}$ only picks up the component $\ket{\Phi^{\lambda}_{\mu}(\mathbf{q})}$ from the mean-field wave function (\ref{eq:HFB_wf}),
   \begin{equation}
  \hat{P}_{\mu\mu}^{\lambda} |\Phi(\mathbf{q})\rangle
  =c_{\lambda,\mu}(\mathbf{q})\ket{\Phi^{\lambda}_{\mu}(\mathbf{q})},
  \end{equation}
  where the coefficient is $c_{\lambda,\mu}(\mathbf{q})$ determined by
   \begin{equation}
   \label{eq:weights}
   |c_{\lambda,\mu}(\mathbf{q})|^2
   =\bra{\Phi(\mathbf{q})} \hat{P}_{\mu\mu}^{\lambda} |\Phi(\mathbf{q})\rangle. 
  \end{equation}
  It defines a normalized symmetry-projected wave function
     \begin{equation}
     \label{eq:projected_wf_normalized}
  \ket{\Phi^{\lambda}_{\mu}(\mathbf{q})}
  =\frac{1}{\sqrt{\bra{\Phi(\mathbf{q})} \hat{P}_{\mu\mu}^{\lambda} |\Phi(\mathbf{q})\rangle}}\hat{P}_{\mu\mu}^{\lambda} |\Phi(\mathbf{q})\rangle. 
  \end{equation}
   The expectation value of a general scalar operator $\hat O$ with respect to this wave function $\ket{\Phi^{\lambda}_{\mu}(\mathbf{q})}$   is determined by
  \begin{eqnarray}
  {\cal O}^{\lambda,\mu}(\mathbf{q},\mathbf{q})
  &=&\frac{ \bra{\Phi^{\lambda}_{\mu}(\mathbf{q})}\hat{O} \ket{\Phi^{\lambda}_{\mu}(\mathbf{q})}}
  {\bra{\Phi^{\lambda}_{\mu}(\mathbf{q})} \mathbb{1} \ket{\Phi^{\lambda}_{\mu}(\mathbf{q})}}
  =\frac{\bra{\Phi(\mathbf{q})}\hat{O}\hat{P}_{\mu\mu}^{\lambda} \ket{\Phi(\mathbf{q})}} 
  {\bra{\Phi(\mathbf{q})} \hat{P}_{\mu\mu}^{\lambda} \ket{\Phi(\mathbf{q})}}.
  \end{eqnarray}
   The kernels can be rewritten explicitly in term of the integration over all the rotation angles $\varphi$, where the integrand is the overlap of  the operator $\hat O$ between the original and rotated mean-field wave functions, weighted by the Wigner-D function, 
     \begin{eqnarray} 
     \label{eq:general_projection_kernel}
      &&\bra{\Phi(\mathbf{q})} \hat O \hat{P}_{\mu\mu}^{\lambda}   \ket{\Phi(\mathbf{q})} \nonumber\\
     &=& \frac{d_{\lambda}}{n_{G}}\int d\mathbf{\varphi} D_{\mu\mu}^{\lambda\ast}(\mathbf{\varphi})  
      \bra{\Phi(\mathbf{q})} \hat O    \ket{\Phi(\mathbf{q};\mathbf{\varphi})}
      \cdot\bra{\Phi(\mathbf{q})}  \hat R(\varphi) \ket{\Phi(\mathbf{q})},
      \end{eqnarray}  
    where the rotated quasiparticle vacuum $\ket{\Phi(\mathbf{q}, g)}$ is defined as below
  \begin{equation}
 \label{eq:rotated_HFB_state}
    \ket{\Phi(\mathbf{q}, \mathbf{\varphi})}\equiv\frac{\hat R(\mathbf{\varphi})\ket{\Phi(\mathbf{q})}}{\bra{\Phi(\mathbf{q})}\hat R(\mathbf{\varphi})\ket{\Phi(\mathbf{q})}}.
  \end{equation}

 \subsection{Some typical examples}

 In general, the HFB wave function $\ket{\Phi(\mathbf{q})}$ may break several symmetries simultaneously. To project out the component with correct quantum numbers, one needs to introduce all the corresponding projection operators. In the following, the projection operators for some typically broken symmetries in $\ket{\Phi(\mathbf{q})}$ will be introduced, separately. 
 
 \subsubsection{The space-reversal symmetry $\mathbb{Z}_2$}
 
 The space-reversal (parity) transformation $\hat{\mathcal P}$ forms the abelian group $\mathbb{Z}_2$ with two elements $\{\hat{\mathbb{1}}, \hat{\mathcal P}\}$. There are two irreps $\hat{\mathcal P}\phi(\mathbf{r}) =\phi(-\mathbf{r})= \pi\phi(\mathbf{r})$, where the eigenvalue $\pi=\pm$ corresponds to even or odd parity, respectively. In atomic nuclei, parity is conserved in the strong interaction and thus it serves as one of the quantum numbers used to label nuclear states.  In atomic nuclei of some particular mass regions with neutron or proton numbers around 34, 56, 88, and 134,  the  negative-parity states of vibration or rotation excitations are observed near their ground states. This phenomenon can be naturally explained in mean-field approaches that the onset of large octupole correlations is attributed to the existence of pairs of single-particle states with opposite parities around the Fermi surface such as $g_{9/2}-p_{3/2}$, $h_{11/2}-d_{5/2}$ and $i_{13/2}-f_{7/2}$, $j_{15/2}-g_{9/2}$ which are strongly coupled by octupole moments~\cite{Butler:1996}.  The emergence of nonzero moments of odd multiplicity such as octupole moments induces the mixing of the states with different parity. As a result, the obtained nuclear wave function $\ket{\Phi({\mathbf{q}})}$ does not have a definite parity. According to (\ref{eq:projected_wf_normalized}), the normalized wave function  with a definite parity $\pi$ can be recovered by projecting out one of the two irreps with a parity-projection operator, 
    \begin{equation}
        \ket{\Phi_\pi(\mathbf{q})}
        = \frac{1}{\sqrt{\bra{\Phi(\mathbf{q})}\hat P^\pi \ket{\Phi(\mathbf{q})}}}\hat P^\pi \ket{\Phi(\mathbf{q})},
    \end{equation}
   where the parity-projection operator $\hat P^\pi$ is defined as
    \begin{equation}
        \hat P^\pi\equiv  \dfrac{1}{2}\Bigg(\mathbb{1} + \pi \hat{\mathcal P}\Bigg), \quad 
        (\hat P^\pi)^2 = \hat P^\pi.
    \end{equation}
   When the space-reversal operator $\hat{\mathcal P}$ acts on the product many-body wave function $\ket{\Phi(\mathbf{q})}$, it induces a factor of $(-1)^\ell$ to the expansion coefficient of each single-nucleon wave function on the spherical HO basis $\varphi_{n\ell jm_j}=R_{n\ell}(r)[Y_{\ell}\otimes \chi_{s}]^{j}_{m_j}$, where $Y_{\ell m}$ and $\chi_{s m_s}$ are the spherical harmonic and spin wave function, respectively. Alternatively,  one can define a many-body operator  $\hat{\mathcal P}=\exp(i\pi{\hat{N}_-})$~\cite{Egido:1991} for the space-reversal operator $\hat{\mathcal P}$, where the operator $\hat{N}_-=\sum_k^\prime c^\dag_k c^{}_k$ is similar to the particle-number operator, but with the summation over all the $k$  restricted to the single-particle states with negative parity only.     The expectation value of a general parity-conserving operator $\hat O$ to the state $ \ket{\Phi_\pi(\mathbf{q})}$ with the definite parity $\pi$ is thus determined by
  \begin{eqnarray}
  {\cal O}^{\pi}(\mathbf{q},\mathbf{q})
  &=&\frac{ \bra{\Phi_\pi(\mathbf{q})}\hat{O} \ket{\Phi_\pi(\mathbf{q})}}
  {\bra{\Phi_\pi(\mathbf{q})} \mathbb{1} \ket{\Phi_\pi(\mathbf{q})}}
  =\frac{\bra{\Phi(\mathbf{q})}\hat{O} \hat P^\pi \ket{\Phi(\mathbf{q})}} 
  {\bra{\Phi(\mathbf{q})}  \hat P^\pi \ket{\Phi(\mathbf{q})}}.
  \end{eqnarray}

 \subsubsection{The gauge symmetry U(1)}

  Nuclear state $\ket{\Phi_{A_\tau}}$ is labeled with a definite number $A_\tau$ of nucleons (neutrons or protons) which is an eigenstate of the rotation operator 
  \begin{equation}
     \hat R (\varphi) \ket{\Phi_{A_\tau}} = e^{-i\varphi A_\tau}\ket{\Phi_{A_\tau}},
 \end{equation}
 where the symbol $A_\tau$ stands for either neutron ($\tau=n$) or proton ($\tau=p$).   This is a {\em local} transformation in the sense that the symmetry transformation is related to {\em pairing vibration} in which nucleon number $A_\tau$ serves as the differentiate coordinate for the dynamics of the pairing vibration in the nuclear system.  The  group element $\hat R (\varphi)$ can be defined in terms of the particle-number operator $\hat A_\tau$ 
 \begin{equation}
     \hat R (\varphi) = e^{i\varphi \hat A_\tau},\quad \hat A_\tau=\sum^{N(Z)}_{j=1} c^\dagger_j c_j,
 \end{equation}
 where $\varphi\in [0, 2\pi]$  is the gauge angle.  This transformation is  called gauge rotation. The symmetry is described by the unitary group acting in one-dimension complex space, and is thus called U(1) gauge symmetry. The wave function $\ket{\Phi_{A_\tau}}$ forms an irrep of the group.
 
 Similar to electrons in a superconducting metal, nucleons inside an atomic nucleus also exhibit pairing correlations, as supported by many evidences including the odd-even staggering of binding energies and the energy gap in the low-lying energy spectra of atomic nuclei~\cite{Bohr:1958}. This pairing correlation can be excellently described with the BCS or HFB theory in terms of a quasiparticle vacuum state $\ket{\Phi(\mathbf{q})}$ which is constructed as an admixture of wave functions with different nucleon numbers. In other words, the obtained wave function $\ket{\Phi(\mathbf{q})}$ is a superposition of different irreps of the U(1) group usually differing from each other by two nucleons. The use of  this particle-number-violating wave function for an atomic nucleus may cause  serious problems as it contains the component of neighbouring atomic nuclei. Some of these problems can be remedied with the help of particle-number projection (PNP) operator $\hat P^{A_\tau}$ which can project out the component $\ket{\Phi_{A_\tau}(\mathbf{q})}$ with the particle number $A_\tau$, 
    \begin{equation} 
        \ket{\Phi_{A_\tau}(\mathbf{q})} 
        = \frac{1}{\sqrt{\bra{\Phi(\mathbf{q})}\hat P^{A_\tau} \ket{\Phi(\mathbf{q})}}}\hat P^{A_\tau} \ket{\Phi(\mathbf{q})},
    \end{equation}
 where the particle-number projection operator is defined according to (\ref{eq:projection-general}),
     \begin{equation}
     \label{eq:PNP_operator}
         \hat P^{{A_\tau}}=  \dfrac{1}{2\pi} \int^{2\pi}_0 d\varphi   e^{-i\varphi  {A_\tau}}\hat R (\varphi).
     \end{equation}
  For atomic nuclei, the numbers of neutrons and protons are conserved separately.  The violation of particle number in the quasiparticle vacuum $\ket{\Phi(\mathbf{q})}$ can be seen clearly by decomposing it into a linear combination of common eigenstates of the particle-number operators $\hat N$ (and $\hat Z$) of neutrons (and protons),  
  \begin{equation}
    \ket{\Phi(\mathbf{q})}=\sum_{N_0Z_0} c_{N_0Z_0}(\mathbf{q}) \ket{\Phi_{N_0Z_0}(\mathbf{q})},\quad 
    \hat N(\hat Z) \ket{\Phi_{N_0Z_0}(\mathbf{q})}=N_0(Z_0) \ket{\Phi_{N_0Z_0}(\mathbf{q})},
 \end{equation}
 where the expansion coefficient $c_{N_0Z_0}$ of the component with $N_0$ neutrons and $Z_0$ protons can be determined by
   \begin{equation}
    \bra{\Phi(\mathbf{q})} \hat P^{N_0}\hat P^{Z_0} \ket{\Phi(\mathbf{q})}
    =\sum_{NZ}  c^\ast_{N_0Z_0}\bra{\Phi_{NZ}(\mathbf{q})}   c_{N_0Z_0}\ket{\Phi_{N_0Z_0}(\mathbf{q})}
    =|c_{N_0Z_0}|^2.
 \end{equation} 
  The expectation value of a general particle-number-conserving operator $\hat O$ to the wave function $\ket{\Phi_{N_0Z_0}(\mathbf{q})}$ is given by
  \begin{eqnarray}
  {\cal O}^{N_0Z_0}(\mathbf{q},\mathbf{q})
  &=&\frac{ \bra{\Phi_{N_0Z_0}(\mathbf{q})}\hat{O} \ket{\Phi_{N_0Z_0}(\mathbf{q})}}
  {\bra{\Phi_{N_0Z_0}(\mathbf{q})} \mathbb{1} \ket{\Phi_{N_0Z_0}(\mathbf{q})}}
  =\frac{\bra{\Phi(\mathbf{q})}\hat{O} \hat P^{N_0}\hat P^{Z_0}\ket{\Phi(\mathbf{q})}} 
  {\bra{\Phi(\mathbf{q})} \hat P^{N_0}\hat P^{Z_0} \ket{\Phi(\mathbf{q})}},
  \end{eqnarray}
   where the kernel can be rewritten explicitly in term of an integration over all the overlaps of  the operator $\hat O$ between the unrotated and rotated mean-field wave functions at different gauge angles ( $\varphi_n, \varphi_p$),
     \begin{eqnarray} 
     \label{eq:PNP_kernel}
     &&\bra{\Phi(\mathbf{q})} \hat O \hat P^{N_0} \hat P^{Z_0} \ket{\Phi(\mathbf{q})}\nonumber\\
     &=&\int^{2\pi}_0 \frac{ e^{-i\varphi_n  N_0}}{2\pi}d\varphi_n
     \int^{2\pi}_0 \frac{ e^{-i\varphi_p  Z_0}}{2\pi}d\varphi_p 
      \bra{\Phi(\mathbf{q})} \hat O  e^{i\hat N  \varphi_n} e^{i\hat Z  \varphi_p} \ket{\Phi(\mathbf{q})}.
      \end{eqnarray}  
If the mean-field wave function $\ket{\Phi(\mathbf{q})}$ has a definite number parity $(-1)^{N_0(Z_0)}$, the interval of the gauge angles can be reduced to be within $[0, \pi]$~\cite{Bally:2021PRC}. The above integration can be carried out using the trapezoidal rule based on the Fomenko expansion method~\cite{Fomenko:1970JPA} in which the particle-number projection operator becomes  
\begin{equation}
\label{eq:Fomenko}
\hat{P}^{A_\tau}=\frac{1}{L} \sum_{m=1}^{L} e^{i\left(\hat{A}_\tau-A_\tau\right) \varphi_m}, \quad \varphi_m=\frac{m}{L}\pi,
\end{equation}
 where the gauge angle $\varphi_\tau\in[0,\pi]$  is discretized with  $L$  points. In atomic nuclei, the wave function $\ket{\Phi(\mathbf{q})}$ is smeared out only a few particle numbers around its mean value, only a small value of $L$ is sufficient to achieve a rather exact projection. Typically, the odd value of $L=7$ or $9$ is employed to avoid the singularity problem of choosing the point $\varphi=\pi/2$~\cite{Tajima:1992NPA,Donau:1998PRC,Anguiano:2001NPA}.

An alternative way to evaluate the integral over the gauge angle $\varphi_\tau$ is to carry out the integration analytically using the {\em residues theorem}, as employed by Dietrich {\em et al.}~\cite{Dietrich:1964}. The projection operator $\hat{P}^{A_\tau}$ in (\ref{eq:PNP_operator}) can be rewritten in terms of a complex variable  $z=e^{i \varphi}$
 \begin{equation}
     \hat{P}^{A_\tau}
     =\frac{1}{2 \pi i} \oint \frac{z^{\hat{A}_\tau}}{z^{A_\tau+1}} dz.
 \end{equation} 

 The projected wave function becomes
 \begin{equation}
  \ket{\Phi_{A_\tau}(\mathbf{q})}
  =\hat{P}^{A_\tau=2p}\ket{\Phi(\mathbf{q})}
  =\frac{1}{2 \pi i} \oint \frac{d \zeta}{\zeta^{p+1}} \prod^\tau_{k=1}\left(u_{k}+v_{k} \zeta a_{k}^{+} a_{k}^{+}\right)|-\rangle,
 \end{equation}
 where the variable $z$ has been replaced  by  $\zeta=z^{2}$. The $p=A_\tau/2$  is the number of pairs. The integration just picks the component with  $\zeta^{-1}$, that is, the component with  $p$  pairs.

The advantage of using the residue theorem is that it provides an analytical way to explore the structural properties of particle-number projected wave functions. Besides, the variation after particle-number projection (PNVAP) can be carried in a similar way to the ordinary variational calculation~\cite{Dietrich:1964}. Compared to the Fomenko expansion, the downside of this method is that it becomes difficult to be implemented in combination with the projection operators of other symmetries, like the rotational symmetry described by the SO(3) group.
 
 \subsubsection{The rotational symmetry SO(3)}
 
 The mean-field wave function $\ket{\Phi(\mathbf{q})}$ for an open-shell nucleus is usually an  admixture of the components $\ket{\Phi_{JM}(\mathbf{q})}$, 
  \begin{equation}
  \label{eq:decomposition_J}
     \ket{\Phi(\mathbf{q})}=\sum_{JM} c_{JM} \ket{\Phi_{JM}(\mathbf{q})},
 \end{equation}
  where $\ket{\Phi_{JM}(\mathbf{q})}$ is the eigenfunction of squared-angular-momentum operator $\hat J^2$ and its projection along $z$-axis $\hat J_z$,
  \begin{subequations}
  \begin{align}
  \hat J^2 \ket{\Phi_{JM}} &= J(J+1)\hbar^2  \ket{\Phi_{JM}},\\ 
  \hat J_z \ket{\Phi_{JM}} &= M\hbar  \ket{\Phi_{JM}}.
 \end{align}
  \end{subequations}
  
  The component  $\ket{\Phi_{JM}(\mathbf{q})}$ can be  obtained by applying an angular-momentum projection (AMP) operator $\hat P^{J}_{MM}$ onto the symmetry-violating wave function.  The AMP operator $\hat P^{J}_{MM}$ can be constructed as a product of two projection operators defined according to the L\"owdin's method~\cite{Lowdin:1964RMP}, 
 \begin{equation}
 \label{eq:Lowdin_projection_Jcomp}
 \hat P^{J}_{MM}=\prod_{I \neq J} \frac{\hat{J}^{2}-I(I+1)}{J(J+1)-I(I+1)}\prod_{K \neq M}\frac{\hat{J}_{z}-K}{M-K}.
 \end{equation}
 From Eqs. (\ref{eq:decomposition_J}) and (\ref{eq:Lowdin_projection_Jcomp}), one immediately finds
  \begin{equation}
  \hat P^{J}_{MM} \ket{\Phi(\mathbf{q})}= c_{JM} \ket{\Phi_{JM}(\mathbf{q})}.
 \end{equation}
  
 Alternatively, one can introduce an operator $\hat P^{J}_{MK}$ with the following form
\begin{equation}
\label{eq:AMP_projection_op_3D}
\hat{P}_{MK}^{J} 
\equiv  
 \dfrac{2J+1}{8\pi^2} \int^{2\pi}_0 d\alpha
     \int^{\pi}_0  \sin\beta d\beta \int^{2\pi}_0 d\gamma
     D^{J\ast}_{MK}(\Omega) \hat R(\Omega),
\end{equation}
where the rotational operator $\hat{R}(\Omega)$ is defined as a rotation about the origin of three-dimensional Euclidean space, specified by three real parameters ($\alpha, \beta, \gamma$)~\cite{Varshalovich:1988}
\begin{equation}
\hat{R}(\Omega) = e^{-i\alpha \hat J_z}e^{-i\beta \hat J_y}e^{-i\gamma\hat J_z}.
\end{equation}
The irrep of  the SO(3) group is the Wigner-D function,
\begin{equation} 
    D_{MK}^{J}(\Omega)=\langle \Phi_{JM}|\hat{R}(\Omega)|\Phi_{JK}\rangle
    =e^{-iM\alpha}d^J_{MK}(\beta)e^{-iK\gamma},
\end{equation}
where the small $d$-function is defined as $d^J_{MK}(\beta)=\bra{\Phi_{JM}} e^{-i\beta \hat J_y}\ket{\Phi_{JM}}$. With the orthogonality relation 
\begin{equation}
     \frac{2J+1}{8\pi^2} \int  d\Omega D_{MK}^{J\ast}(\Omega)   D^{J'}_{K'M'}(\Omega)
     =\delta_{JJ'} \delta_{MK'}\delta_{KM'},
\end{equation} 
one can prove that the operator (\ref{eq:AMP_projection_op_3D}) extracts the component $\ket{\Phi_{JM}}$ from the symmetry-breaking wave function $\ket{\Phi}$ with the weight $c_{JK}$,
\begin{eqnarray}
\label{eq:AMP}
\hat{P}_{MK}^{J} \ket{\Phi(\mathbf{q})} 
&=& c_{JK} |\Phi_{JM}\rangle.
\end{eqnarray}
Thus, the operator $\hat{P}_{MK}^{J}$ can also be written as
 \begin{equation}
    \hat{P}_{MK}^{J} = \sum_\sigma |\Phi^\sigma_{JM}\rangle \langle \Phi^\sigma_{JK}|,
\end{equation} 
where $\sigma$ characterizes all quantum numbers besides the $J, M$ required to completely specify the states. From the above one arrives at the following relation 
 \begin{equation}
    \hat{P}_{MK}^{J} \hat{P}_{M'K'}^{J'} = \delta_{JJ'}\delta_{KM'} \hat{P}_{MK'}^{J},\quad 
    \hat{P}_{MK}^{J\dagger}= \hat{P}_{KM}^{J}.
 \end{equation}

 In other words, the operator $\hat{P}_{MK}^{J}$ is a true projection operator only if  $M=K$. In a general case, the operator $\hat{P}_{MK}^{J}$ with $M\neq K$ is sometimes referred to as {\em transfer operator}. See such as Ref.~\cite{Bally:2021PRC} and reference therein for more discussions.

The wave function of a triaxially deformed nucleus ($\beta_{22}\neq 0$) with the good angular momentum $J$ can be constructed as~\cite{Yao:2008CPL,Yao:2009PRC}
 \begin{equation}
 \label{eq:triaxial_wf}
\ket{\Phi_{JM}(\mathbf{q})}
=\sum_{K} g^J_{K} \hat{P}_{M K}^{J}|\Phi(\mathbf{q})\rangle.
\end{equation} 
In the simple case that the wave function  $|\Phi\rangle$  has axial symmetry,  the $K$ value is conserved. Otherwise, one needs to mix the components with different $K$ values. The mixing coefficient $g^J_{K}$ is determined by minimizing the energy of state which is given by
 \begin{equation}
{\cal E}^{J}(\mathbf{q},\mathbf{q})
=\frac{\left\langle\Phi_{J M}(\mathbf{q}) |\hat H_0| \Phi_{J M}(\mathbf{q})\right\rangle}{\left\langle\Phi_{J M} (\mathbf{q})\mid \Phi_{J M}(\mathbf{q})\right\rangle}
=\frac{\sum_{K K^{\prime}} g_{K}^{J*} g^J_{K^{\prime}}  H_{K K^{\prime}}^{J}}
{\sum_{K K^{\prime}} g_{K}^{I*} g^J_{K^{\prime}} N_{K K^{\prime}}^{J}}
\end{equation} 
where the relation $\hat P^J_{KM}\hat H_0\hat P^J_{MK'}=\hat H_0\hat P^J_{KK'}$ is used. The Hamiltonian and norm kernels are defined below
\begin{subequations}
 \begin{align}
H_{K K^{\prime}}^{J}&=\left\langle\Phi(\mathbf{q})\left|\hat H_0 \hat{P}_{K K^{\prime}}^{J}\right| \Phi(\mathbf{q})\right\rangle \\
N_{K K^{\prime}}^{J}&=\left\langle\Phi(\mathbf{q})\left|\hat{P}_{K K^{\prime}}^{J}\right| \Phi(\mathbf{q})\right\rangle.
\end{align}
\end{subequations}
The coefficient  $g^J_{K}$ is the solution to the generalized eigenvalue problem
\begin{equation} 
\sum_{K^{\prime}} H_{K K^{\prime}}^{J} g^J_{K^{\prime}}={\cal E}^{J} \sum_{K^{\prime}} N_{K K^{\prime}}^{J} g^J_{K^{\prime}}.
\end{equation} 
This equation is equivalent to the diagonalization of the Hamiltonian in the space spanned by the nonorthogonal basis functions $\hat{P}_{M K}^{J}|\Phi(\mathbf{q})\rangle$.

Before ending this subsection, it is worth mentioning that the deformed state $\ket{\Phi(\mathbf{q})}$ becomes a stationary state if the atomic nucleus is very large~\cite{Blaizot:1986}. Considering the nuclear system evolves with time $t$ from the deformed state $\ket{\Phi(\mathbf{q})}$, the wave function at the time $t$ is given by
  \begin{equation}
  \label{eq:time_dependent_deformed_state}
      \ket{\Psi(t)} = e^{-i\hat H_0 t/\hbar}\ket{\Phi(\mathbf{q})}
      =e^{-iE_0t/\hbar} \sum_{JM} c_{JM} e^{-iJ(J+1)t/2{\cal J}}\ket{\Phi_{JM}(\mathbf{q})} 
  \end{equation}
  where the $\ket{\Phi_{JM}(\mathbf{q})}$ is assumed to be the eigenstate of the Hamiltonian $\hat H_0$ (independent of time) with the eigenvalue $E_J$ taking the following value,
    \begin{equation}
      \hat H_0 \ket{\Phi_{JM}(\mathbf{q})} 
      = E_J\ket{\Phi_{JM}(\mathbf{q})},\quad E_J=E_0+J(J+1)/2{\cal J}.
  \end{equation} 
  In the limit that the moment of inertial ${\cal J}\to\infty$  for a very large nuclear system, the wave function in  (\ref{eq:time_dependent_deformed_state}) becomes  
  \begin{equation}
      \ket{\Psi(t)}_{{\cal J}\to\infty} = 
      e^{-iE_0t/\hbar} \sum_{JM} c_{JM} \ket{\Phi_{JM}(\mathbf{q})} 
      = e^{-iE_0t/\hbar} \ket{\Phi(\mathbf{q})}.
  \end{equation}
  The above formulas indicate that {\em the deformed state $\ket{\Phi(\mathbf{q})}$ from mean-field calculations is approximately a stationary state for a very large nucleus.}  In other words, for an infinite large system, the energy splitting between different $J$ states vanishes, the direct couplings between any two rotated symmetry-breaking states $\ket{\Phi(\mathbf{q}, \Omega_1)}\equiv\hat R(\Omega_1)\ket{\Phi(\mathbf{q})}$  and $\ket{\Phi(\mathbf{q}, \Omega_2)}$ with $\Omega_1\neq \Omega_2$ are zero, i.e.,
  \begin{equation}
      \bra{\Phi(\mathbf{q}, \Omega_1)} \Phi(\mathbf{q}, \Omega_2)\rangle=\delta(\Omega_1-\Omega_2).
  \end{equation}
  It means that the time required for the nuclear system to pass from one orientated  state to another orientated state becomes infinite long. It explains why the restoration of rotational symmetry is more important in light deformed nuclear systems than in heavy ones.

\subsubsection{Isospin symmetry SU(2)}

As discussed before, the isospin symmetry is a dynamical symmetry in atomic nuclei, described by the special unitary group acting in two-dimension complex space, namely, the SU(2) group. It plays an important role in understanding the asymmetry between the energy spectra of mirror nuclei and nuclear $\beta$ decays. The  isospin symmetry is largely preserved by the strong interactions with a small violation from weakly {\em charge-dependent} components and the mass difference between neutrons and protons. Besides, the main source of isospin breaking is the electromagnetic interaction. For the convenience of discussion, the total nuclear Hamiltonian 
\begin{equation}
\label{eq:isospin-H}
    \hat H = \hat H_0 + \hat H_1,
\end{equation}
is separated into a sum of isospin-rotation invariant term $\hat H_0$ 
and isospin-rotation breaking term $\hat H_1$  dominated by the Coulomb interaction $V_C$
\begin{equation}
\label{eq:isospin-H-commutate}
    \quad [\hat H_0, \hat R_y(\theta)]=0,\quad [\hat H_1, \hat R_y(\theta)] \neq 0,
\end{equation}
 where the  operator $\hat{R}_y(\theta)=e^{-i \theta \hat{T}_{y}}$   is defined as a rotation about the  $y$-axis in the isospin space with the angle $\theta$ with the  $\hat T_y$ being the $y$-component of isospin operator.  With the $\hat H_0$ term only,  the Hamiltonian commutes with the total isospin $\hat T^2$. Nuclear states  can then be labeled with the quantum number $T$, and the states $\ket{TM_T}$ form degenerate multiplets consisting of $2T+1$ components with different $M_T$. With the $V_C$, the degeneracy between the multiplet components is lifted. This phenomenon is shown to be a general feature in the spectra of isobaric isotopes.
 
 In addition to the $\hat H_1$ term which breaks the isospin symmetry explicitly, in the mean-field description of atomic nuclei with $N\neq Z$,  the presence of the neutron or proton excess automatically yields isovector mean fields, i.e., different mean-field potentials for protons and neutrons. As a result, the isospin invariance is artificially broken by the mean-field approximation even in the case that the isospin-invariant Hamiltonian is used~\cite{Brink:1970,Engelbrecht:1970PRL}. It indicates that more efforts are needed to take care of the isospin  mixing in atomic nuclei (especially the neutron-rich nuclei with large neutron excess) if one starts from the mean-field approximation.  Starting from an isospin-invariant Hamiltonain $\hat H_0$, the breaking of isospin invariance in the mean-field solution is then due to the omission of neutron-proton correlations and can be recovered with the random-phase approximation (RPA)~\cite{Engelbrecht:1970PRL}. From this point of view, the RPA provides an excellent framework to consider isospin-breaking effect in atomic nuclei, which is essential to achieve an accurate calculation of the isospin corrections for superallowed Fermi beta decay~\cite{Liang:2009}.   
 
 Alternatively, one can also take into account the isospin mixing effects with the isospin-symmetry projection operator~\cite{Caurier:1980PLB,Satula:2009PRL}. This projection method applies not only to the ground state but also to nuclear excited states. In this framework, the nuclear state $\ket{n, M_T}$ with a proper isospin mixing can be spanned in terms of the irreps $\ket{TM_T}$ of the SU(2) group, 
\begin{equation}
    \ket{n, M_T}=\sum_{T \geq\left|M_T\right|} a_{TM_T}^{n}\ket{TM_T}
\end{equation}
where the $\ket{TM_T}$ is  generated by acting the isospin-projection operator $\hat{P}_{M_TM_T}^{T}$  on top of the mean-field wave function $|\Phi_{N_0Z_0}\rangle$~\footnote{Note that a three-dimension isospin projection operator is needed if a particle-number-violating mean-field wave function $\ket{\Phi(\mathbf{q})}$ is employed.} which is chosen as the eigenstate  of neutron- and proton-number operators~\cite{Satula:2009PRL,Satula:2010PRC}
\begin{align}
\left|T M_T\right\rangle &=\frac{1}{\sqrt{\langle\Phi_{N_0Z_0}|\hat{P}_{M_TM_T}^{T}|\Phi_{N_0Z_0}\rangle}} \hat{P}_{M_TM_T}^{T}|\Phi_{N_0Z_0}\rangle \nonumber\\
&=\frac{1}{\sqrt{\langle\Phi_{N_0Z_0}|\hat{P}_{M_TM_T}^{T}|\Phi_{N_0Z_0}\rangle}}\frac{2 T+1}{2}  \int_{0}^{\pi} d \theta \sin \theta d_{M_TM_T}^{T}(\theta) \hat{R}_y(\theta)|\Phi_{N_0Z_0}\rangle,
\end{align}
where  $d_{T_{z} T_{z}}^{T}(\beta)$  is the Wigner $d$-function, and  $M_T=(N_0-Z_0) / 2$  is the third component of the total isospin  $T$.  The coefficient $a_{T M_T}^{n}$ defining the degree of isospin mixing is determined by the variational principles using the full Hamiltonian in Eq.(\ref{eq:isospin-H}). Usually, an isospin-mixing parameter $\alpha_C$ is introduced. For the lowest-energy solution with $n=1$,  it is defined as  $\alpha_{C}=1-\left|a_{TM_T}^{n=1}\right|^{2}$ with $T=M_T$.

   \subsection{Approximate treatments with power expansions}

   Extracting  the  irreps of a symmetry group  of a given nuclear Hamiltonian with the projection operator usually requires a multi-dimensional integration of many overlaps between two general Slater determinants for all the group elements. The calculation of these overlaps is  generally very time-consuming. This is particularly true if the VAP scheme is employed where this kind of integration needs to be carried out at each iteration.   Historically, some approximate treatments of the symmetry-restoration effect on nuclear energy have been proposed, such as the Lipkin~\cite{Lipkin:1960} and Kamlah~\cite{Kamlah:1968} power expansion methods. 
   
   The basic idea of these expansion methods is to eliminate the symmetry-violation effect on the energy eigenvalue by introducing a model Hamiltonian~\cite{Lipkin:1960},   
   \begin{equation}
       H^\prime  = \hat H_0 - f(\hat X),
    \end{equation} 
    for which all the states $\ket{\Phi_{X_0}}$ belonging to different irreps (labelled with $X_0$) of the symmetry group $G$ are degenerate.  Here, $\hat H_0$ is the original Hamiltonian, and the $f(\hat X)$ is chosen as a power series expansion of the operator $\hat X$,
    \begin{equation}
    \label{eq:Lipkin_expansion}
       f(\hat X) = \sum^M_{k=1}\lambda_k (\hat X -X_0)^k, \quad \hat X\ket{\Phi_{X_0}} = X_0\ket{\Phi_{X_0}},
    \end{equation}
    where the operator $\hat X$ is given by the generators of the symmetry group and can be chosen as for instance the angular-momentum operator $\hat J^2$ or particle-number operator $\hat N$ if  the rotational SO(3) symmetry or the gauge U(1) symmetry is violated, respectively.  To this end, the expansion coefficients $\lambda_k$ are determined in such a way that the quantity $\bra{\Phi_{X_0}}\hat{H}'\ket{\Phi_{X_0}}$ is independent of the $X_0$. 
   
    The Lipkin's idea was adopted by Nogami~\cite{Nogami:1964} to suppress the impact of particle-number fluctuation of the BCS wave function on the binding energy of open-shell atomic nuclei  by truncating the expansion in (\ref{eq:Lipkin_expansion}) up to the $k=2$ terms,
       \begin{equation}
    \label{eq:Nogami_expansion}
       f(\hat N) = \lambda_1(\hat N - N_0)+ \lambda_2(\hat N^2-N^2_0).
    \end{equation}
    The above prescription is referred to as the Lipkin-Nogami method.
   
   The Lipkin method can be alternatively understood as an expansion of the energy of the symmetry-projected wave function, which leads to the Kamlah's method~\cite{Kamlah:1968}, where a general scheme to approximate the energy of momentum projection was derived, and it was shown that the cranking model is simply the variational problem in the first-order approximation to the energy of the angular-momentum projected state. Following Ref.~\cite{Ring:1980}, this method is introduced for the case of a one-dimensional rotation $\hat R(\varphi)=e^{i\varphi \hat X}$, where the parameter $\varphi$ can be either the Euler angle  associated with angular momentum $J$~\footnote{As shown in (\ref{eq:AMP_projection_op_3D}), a three-dimensional rotation operator should be employed in principle to defined the angular-momentum projection. The Kamlah expansion for the general three-dimensional rotation case was presented by Beck, Mang, and Ring~\cite{Beck:1970}.   } or the gauge angle associated with particle number $N$. The energy of the projected state labeled with $X_0$ is generally written as 
   \begin{equation}
   {\cal E}^{X_0}(\mathbf{q}, \mathbf{q})
   =\frac{\int d \varphi e^{-i \varphi X_0}\left\langle\Phi(\mathbf{q})\left|\hat{H}_0 e^{i \varphi \hat{X}}\right| \Phi(\mathbf{q})\right\rangle}{\int  d \varphi e^{-i \varphi X_0}\left\langle\Phi(\mathbf{q})\left|e^{i \varphi \hat{X}}\right| \Phi(\mathbf{q})\right\rangle}
   =\frac{\int  d \varphi e^{-i \varphi X_0} h(\varphi)}{\int d \varphi e^{-i \varphi X_0} n(\varphi)}
   \end{equation} 
   where the short-hand notations $h(\varphi)$ and $n(\varphi)$ were introduced for Hamiltonian  and norm overlaps, respectively
   \begin{equation}
   \label{eq:overlap_functions}
       h(\varphi) =\left\langle\Phi(\mathbf{q})\left|\hat{H}_0 e^{i \varphi \hat{X}}\right| \Phi(\mathbf{q})\right\rangle,\quad n(\varphi) =   \left\langle\Phi(\mathbf{q})\left|e^{i \varphi \hat{X}}\right| \Phi(\mathbf{q})\right\rangle.
   \end{equation}
   In the case that the deformation or pairing correlations are  strong in $\ket{\Phi(\mathbf{q})}$, one expects the  $h(\varphi)$  and  $n(\varphi)$ to be peaked at  $\varphi=0$  and to be very small elsewhere in such a way that the quotient  $h(\varphi) / n(\varphi)$  behaves smoothly. In this case, one can expand  $h(\varphi)$  in terms of  $n(\varphi)$  in the following way
     \begin{equation}
     \label{eq:energy_overlap_expansion}
     h(\varphi)=\sum_{m=0}^{M} h_{m} \hat{\mathcal{K}}^{m} n(\varphi),\quad \hat{\mathcal{K}}=\frac{1}{i} \frac{\partial}{\partial \varphi}-\langle\Phi|\hat{X}| \Phi\rangle,
  \end{equation} 
  where the {\em Kamlah operator} $\hat{\mathcal{K}}$ was introduced. In this case, the energy of the projected state becomes
   \begin{eqnarray}
   \label{eq:projected_energy_expansion}
   {\cal E}^{X_0}_{[M]}
   &=&
   \frac{\int  d \varphi e^{-i \varphi X} \sum_{m=0}^{M} h_{m} \hat{\mathcal{K}}^{m} n(\varphi)}
   {\int  d \varphi e^{-i \varphi X} n(\varphi)}
   =\sum^M_{m=0} h_m \left(\bra{\Phi}(\Delta \hat X)\ket{\Phi} \right)^m,
   \end{eqnarray}
    with $\Delta \hat X\equiv \hat X- \bra{\Phi} \hat X\ket{\Phi}$.   The expansion coefficients  $h_{m}$  are determined by applying the Kamlah operators  $1$, $\hat{\mathcal{K}}$, $\ldots$, $\hat{\mathcal{K}}^{M}$ and taking the limit $\varphi\to0$. One has the following relation
   \begin{equation}
   \label{eq:LN_shift}
  \bra{\Phi}\hat H(\Delta \hat X)^n\ket{\Phi}
  = \sum^M_{m=0} h_m \bra{\Phi}(\Delta \hat X)^{m+n} \ket{\Phi}.
   \end{equation} 
   Different levels of approximations can be defined according to the truncation in the expansion order $M$.
   \begin{itemize}
       \item Up to the first order $(M=1)$: In this case, the Eq.(\ref{eq:LN_shift}) produces two equations,
       \begin{subequations}
   \begin{align}
     \bra{\Phi}\hat H \ket{\Phi}
     &=   h_0 + h_1 \bra{\Phi}(\Delta \hat X)  \ket{\Phi}  \\  
     \bra{\Phi}\hat H(\Delta \hat X)\ket{\Phi}
    &= h_0 \bra{\Phi}(\Delta \hat X) \ket{\Phi} + h_1   \bra{\Phi}(\Delta \hat X)^2 \ket{\Phi}.
   \end{align} 
   \end{subequations}
   The projected energy in (\ref{eq:projected_energy_expansion}) is composed of two terms,
    \begin{equation}
       {\cal E}^{X_0}_{[1]}
       =h_0+ h_1\bra{\Phi}(\Delta \hat X)  \ket{\Phi}.
    \end{equation}
    The wave function $\ket{\Phi}$ is usually determined by the minimization of the projected energy ${\cal E}^{X_0}_{[1]}$  with the constraint $\bra{\Phi}(\Delta \hat X)  \ket{\Phi}=0$ which in the case of $\hat X=\hat J$ leads to the equation in the cranking model~\cite{Ring:1980}
      \begin{equation}
        \frac{\delta}{\delta \Phi}
        \Bigg(\langle\Phi|\hat H_0| \Phi\rangle
        -h_1 \langle\Phi|\hat{X}| \Phi\rangle\Bigg)=0,\quad h_1=\frac{\bra{\Phi}\hat H_0(\Delta \hat X)\ket{\Phi}}{\bra{\Phi}(\Delta \hat X)^2 \ket{\Phi}},
         \end{equation} 
    where $h_1$ is referred to as the cranking frequency.      
  \item  Up to the second order $(M=2)$:  In this case, the above relation leads to a set of equations corresponding to the choices of $n=0, 1, 2$, respectively,
\begin{subequations}
   \begin{align}
     \bra{\Phi}\hat H_0 \ket{\Phi}
     &=   h_0 + h_1 \bra{\Phi}(\Delta \hat X)  \ket{\Phi} + h_2  \bra{\Phi}(\Delta \hat X)^2 \ket{\Phi}\\  
     \bra{\Phi}\hat H_0(\Delta \hat X)\ket{\Phi}
    &= \sum^2_{m=0} h_m \bra{\Phi}(\Delta \hat X)^{m+1} \ket{\Phi},\\ 
      \bra{\Phi}\hat H_0(\Delta \hat X)^2\ket{\Phi}
    &= \sum^2_{m=0} h_m \bra{\Phi}(\Delta \hat X)^{m+2} \ket{\Phi}.
   \end{align} 
   \end{subequations}
  One can determine the coefficients $h_0, h_1$ and $h_2$ from the above equations with the constraint $\bra{\Phi}(\Delta \hat X)  \ket{\Phi}=0$ and the omission of the $\bra{\Phi}(\Delta \hat X)^3 \ket{\Phi}$ term,  
  \begin{subequations}
   \begin{align}
     h_0 &= \bra{\Phi}\hat H_0 \ket{\Phi} - h_2  \bra{\Phi}(\Delta \hat X)^2 \ket{\Phi},\quad
    h_1  = \frac{\bra{\Phi}\hat H_0(\Delta \hat X)\ket{\Phi}}{\bra{\Phi}(\Delta \hat X)^2 \ket{\Phi}},\\ 
     h_2 & =\frac{\bra{\Phi}\hat H_0(\Delta \hat X)^2\ket{\Phi}- \bra{\Phi}  \hat H_0 \ket{\Phi} \bra{\Phi}(\Delta \hat X)^2 \ket{\Phi}}{\bra{\Phi}(\Delta \hat X)^4 \ket{\Phi}-\bra{\Phi}(\Delta \hat X)^2 \ket{\Phi}^2},
   \end{align} 
   \end{subequations}
 from which one obtains the energy of the projected state 
   \begin{equation}
  {\cal E}^{X_0}_{[2]} = h_0 + h_1 \bra{\Phi}(\Delta \hat X)\ket{\Phi} + h_2 \bra{\Phi}(\Delta \hat X)^2 \ket{\Phi}.
   \end{equation}
   The wave function $\ket{\Phi}$ is determined by minimizing the energy ${\cal E}^{X_0}_{[2]}$. For the case of $\hat X=\hat N$, the truncation at this order is nothing but the formulas proposed by Nogami~\cite{Nogami:1964}. For simplicity, the $h_2$ is often kept to be constant and is determined after the variation. 
  
   \end{itemize}
   As mentioned before, the truncation of the Kamlah expansion  in (\ref{eq:energy_overlap_expansion}) converges rather rapidly for the atomic nuclei with strong correlations for which case the overlap functions    $h(\varphi)$  and  $n(\varphi)$ are strongly peaked at the origin and the fluctuation $\bra{\Phi} (\Delta \hat X)^2\ket{\Phi}$ is very large. For example, the wave function $\ket{\Phi}$  with a large fluctuation in either angular momentum $\bra{\Phi} (\Delta \hat J)^2\ket{\Phi}$ or particle number $\bra{\Phi} (\Delta \hat N)^2\ket{\Phi}$ indicating the spread $\bra{\Phi}\Delta\varphi^2\ket{\Phi}$ in the corresponding Euler angle or gauge angle is small. In other words, the orientation of the nucleus is almost fixed in space.  For the nuclei with weak collective correlations (small fluctuation),  however, one may need to include higher-order terms in the  expansion~\cite{WangXB:2014}.  
   
   Finally, it is worth noting that these approximate methods only provide a correction to energy. The resultant wave function $\ket{\Phi(\mathbf{q})}$ is still an admixture of different irreps of the symmetry group and thus the  quantum number $X_0$ is not recovered. One still needs to implement the projection operator to extract the component $\ket{\Phi_{X_0}(\mathbf{q})}$ with correct quantum numbers from the symmetry-breaking wave function $\ket{\Phi(\mathbf{q})}$ to study (electromagnetic multiple) transitions between different states.

 \subsection{Electric multiple transitions}
 
 Symmetry restoration plays an essential role in the studies of transition between nuclear states with different spin-parity quantum numbers. It is also one of the main motivations to implement symmetry-restoration methods into symmetry-breaking calculations. With the symmetry-projected wave functions $\ket{\Phi_{J_{i/f},\sigma_{i/f}}}$ of the initial and find states defined in (\ref{eq:triaxial_wf}), where $\sigma$ distinguishes different states for the same angular momentum $J$, one can evaluate the transition strength between these two states under a general tensor operator $\hat{Q}_{\lambda \mu}$  using the  {\em Wigner-Eckart} theorem,
  \begin{equation}
  \label{eq:transition_strengths}
   B(E\lambda; J_i\sigma_i \to J_f\sigma_f)
   =\frac{e^2}{2J_i+1} \Bigg|\left\langle \Phi_{J_{f},\sigma_f}(\mathbf{q}_{f})\left\|\hat{Q}_{\lambda}\right\| \Phi_{J_{i},\sigma_i}(\mathbf{q}_{i})\right\rangle\Bigg|^2,
 \end{equation}
 where the reduced transition matrix element is determined by
\begin{align}
  \label{eq:reduced_matrix_elements}
&\left\langle \Phi_{J_{f},\sigma_f}(\mathbf{q}_{f})\left\|\hat{Q}_{\lambda}\right\| \Phi_{J_{i},\sigma_i}(\mathbf{q}_{i})\right\rangle \nonumber\\
&=(2{J}_{f}+1) \sum_{K_{i} K_{f}} g_{K_{f},\sigma_{f}}^{J_{f}\ast}  
g_{K_i,\sigma_{i}}^{J_{i}} 
\sum_{\mu K^{\prime}}(-1)^{J_{f}-K_{f}+\lambda}\left(\begin{array}{ccc}
J_{f} & \lambda & J_{i} \\
-K_{f} & \mu & K^{\prime}
\end{array}\right)\nonumber\\
&\times\left\langle\Phi\left(\mathbf{q}_{f}\right)\left|\hat{Q}_{\lambda \mu} \hat{P}_{K^{\prime} K_{i}}^{J_{i}}\right| \Phi\left(\mathbf{q}_{i}\right)\right\rangle.
\end{align} 
In the above derivation, the following relation is used, 
 \begin{equation}
  \hat{P}_{K M}^{J} \hat{Q}_{\lambda \mu} \hat{P}_{M^{\prime} K^{\prime}}^{J^{\prime}}
  =   \langle J^{\prime} M^{\prime} \lambda \mu| JM\rangle
  \sum_{\nu \mu^{\prime}}  
    \langle J^{\prime} \nu \lambda \mu^{\prime} | JK\rangle
  \hat{Q}_{\lambda \mu^{\prime}} \hat{P}_{\nu K^{\prime}}^{J^{\prime}},
 \end{equation} 
 with $\langle J^{\prime} M^{\prime} \lambda \mu| JM\rangle$ being a Clebsch-Gordan coefficient. 
 
 It is noted that in Eq.(\ref{eq:reduced_matrix_elements}) the mean-field wave functions of the initial and final states are characterized with different collective coordinates $\mathbf{q}_i$ and $\mathbf{q}_f$. In the PGCM, these collective coordinates are also integrated out weighted by the functions $f_{J_i, \sigma_i}(\mathbf{q}_i)$ and $f_{J_f, \sigma_f}(\mathbf{q}_f)$, respectively, as shown in Eq.(\ref{eq:GCM_wf}).

\section{Some illustrative applications}

%  There are different computational programs performing symmetry-restoration calculations based on different effective interactions or EDFs for atomic nuclei. Only a few of them are published, such as HFODD~\cite{Schunck:2016CPC}, TAURUS~\cite{Bally:2021EPJA}, and HFBTHO~\cite{Marevic:2022CPC}.
 
%  In this section, we will introduce some illustrative applications of the symmetry-restored methods, including 
 
 \subsection{Impact of particle-number projection in low-lying states}
 
\begin{figure}[tb]
\centering
\includegraphics[width=\textwidth]{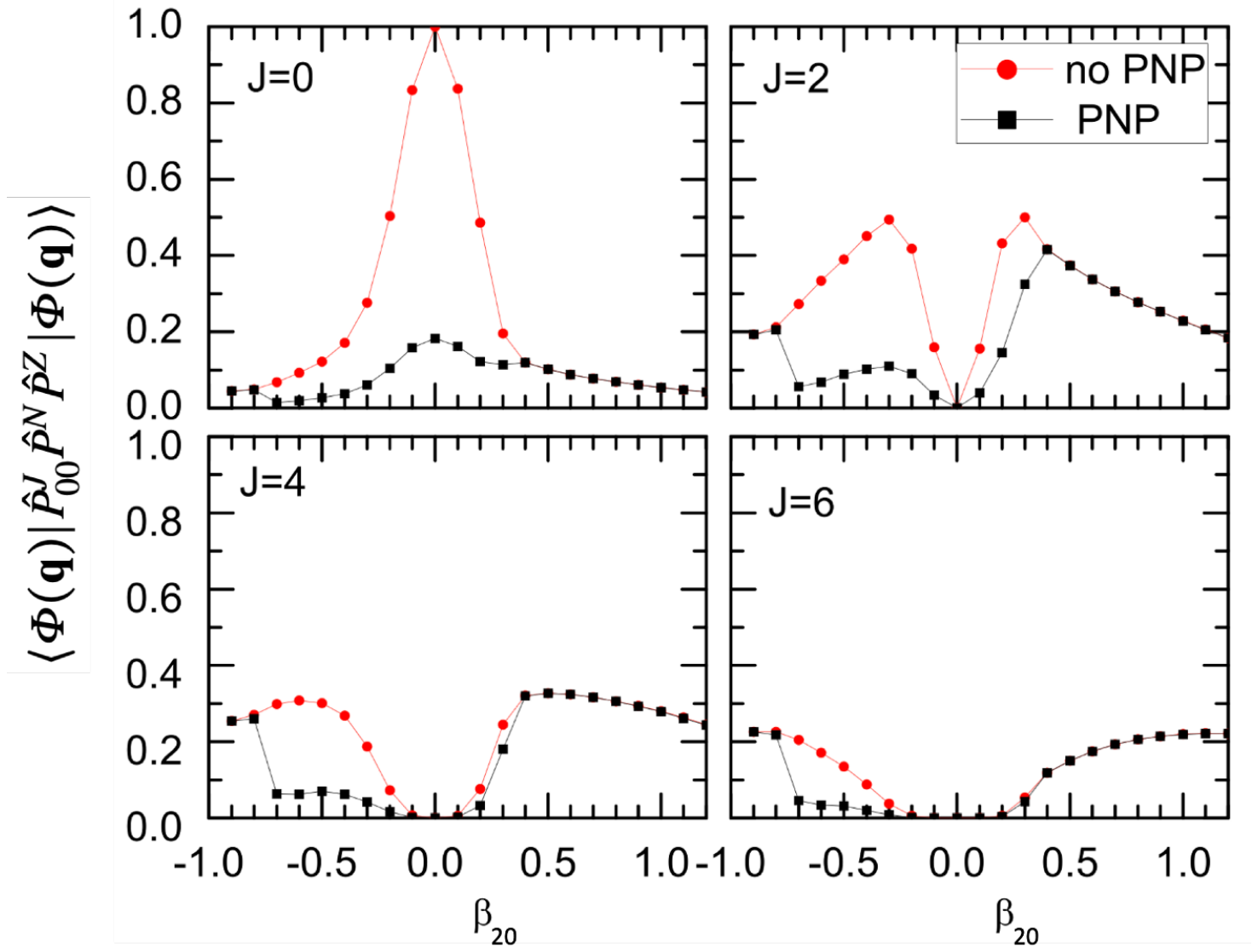}
\caption{The  norm kernel $\bra{\Phi(\mathbf{q})}\hat P^J_{00}\hat P^N\hat P^Z\ket{\Phi(\mathbf{q})}$ as a function of the quadrupole deformation parameter $\beta_{20}$  for  $^{24}$Mg, where the intrinsic wave functions $\ket{\Phi(\beta_{20}}$ are from the constrained CDFT calculations. The number $L$ of meshes in gauge angles is set to either 1 or 7, labelled with ``no PNP" and ``with PNP", respectively. 
}
\label{fig:Mg24_Norm}
\end{figure}

 Figure~\ref{fig:Mg24_Norm} displays the norm kernel $\bra{\Phi(\mathbf{q})}\hat P^J_{00}\hat P^{Z_0}\hat P^{N_0}\ket{\Phi(\mathbf{q})}$ of mean-field state $\ket{\Phi(\mathbf{q})}$ of $^{24}$Mg  with projection onto angular momentum $J=0, 2, 4, 6$,  neutron number $N_0=12$ and proton number $Z_0=12$ as a function of quadrupole deformation parameter $\beta_{20}$ from the calculation of a covariant density functional theory (CDFT)~\cite{Yao:2011} with the PC-PK1~\cite{Zhao:2010PRC} force. The norm kernels without the particle-number projection are also shown for comparison. According to (\ref{eq:weights}),  the norm kernel determines the weight  $|c_{JN_0Z_0}(\mathbf{q})|^2$ of the component $\ket{\Phi_{JN_0Z_0}(\mathbf{q})}$ in the mean-field wave function $\ket{\Phi(\mathbf{q})}$. It is shown clearly that  the spherical mean-field state with $\beta_{20}=0$ contains only the $J=0$ component. Besides, one notices that the particle-number projection does not change the kernels of prolate deformed states with $\beta_{20}\ge0.4$, indicating that particle numbers are not violated in these states. It is because of the collapse of pairing correlations in these states. In contrast, pairing correlation is shown to be strong in oblate deformed states with the deformation parameter $\beta_{20}\simeq -0.3$, leading to a strong violation of particle numbers. As a result, these oblate states contain the component with particle numbers other than $N_0=Z_0=12$ and the weight $|c_{JN_0Z_0}(\mathbf{q})|^2$ is significantly decreased when the particle-number projection is turned on. 
 
\begin{figure}[] 
\includegraphics[width=0.9\textwidth]{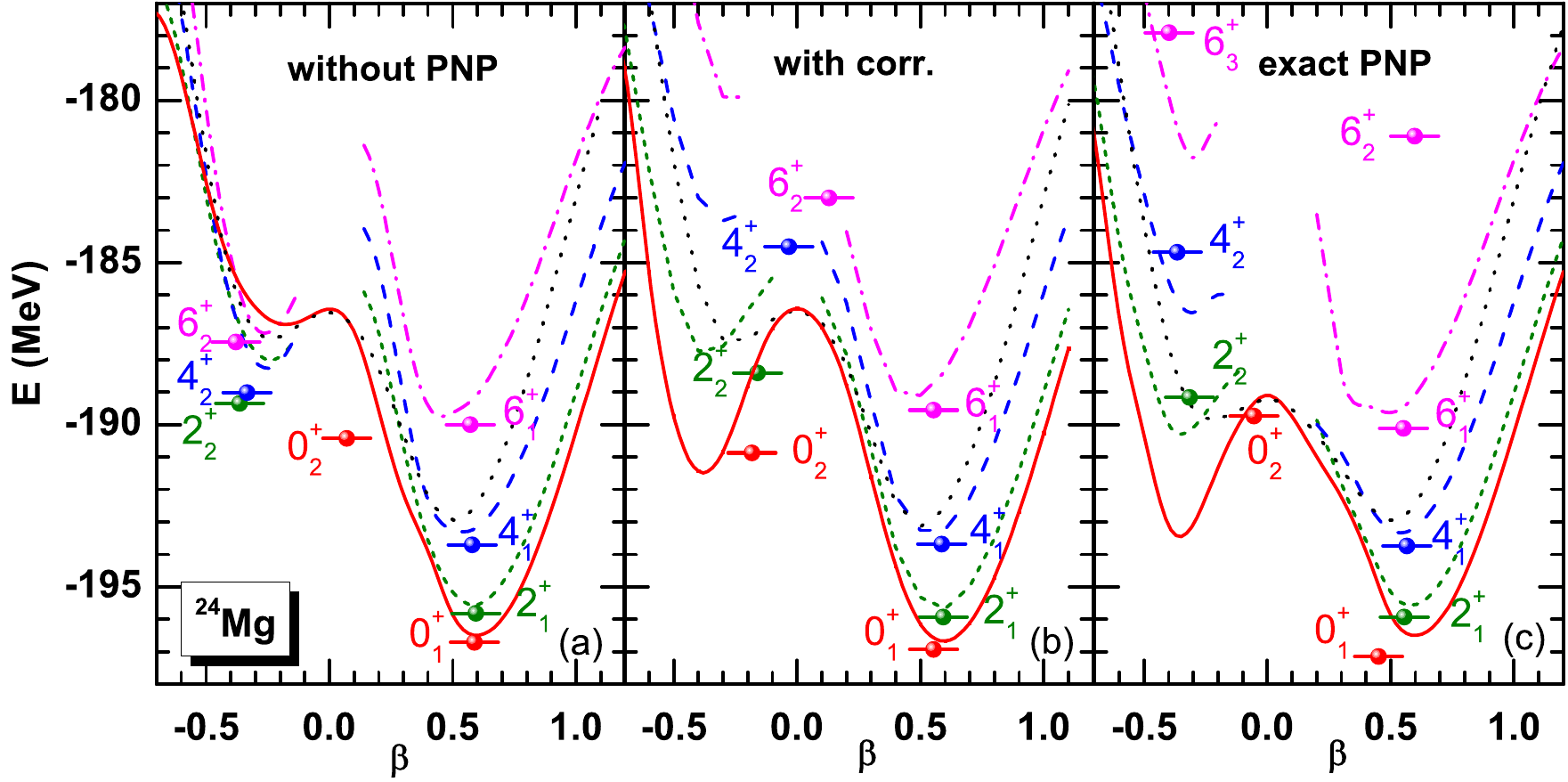}
\caption{Total energy curves projected onto angular momentum ($J=0, 2, \cdots, 6$), as well as the projected GCM states in $^{24}$Mg for the cases of (a) without PNP, (b) with an approximate particle-number correction (\ref{eq:Fermi_energy_correction}), and (c) with exact PNP. The projected energy curves are plotted as a function of the intrinsic deformation $\beta_{20}$ of the mean-field states. The energies of projected GCM states are indicated by bullets and horizontal bars placed at the average deformation $\bar\beta_{J,\sigma}$ defined in (\ref{eq:average_beta_GCM}).
}
\label{fig:Mg24_PNP}
\end{figure}

Starting from each mean-field state $\ket{\Phi(\mathbf{q})}$, one can evaluate the energy of the symmetry-projected state $\ket{\Phi_{JN_0Z_0}(\mathbf{q})}$ for different value of $J$, and it is usually called {\em projected energy surface}. The results of different spin states are shown in Fig.~\ref{fig:Mg24_PNP}, where the violation of particle numbers is treated in three different ways. The panels (a) and (c) show the results without and with the particle-number projection, respectively, while the panel (b) shows the approximate way to consider the symmetry-violation effect on the projected energy, in which the original Hamiltonian $H_0$ is replaced with the following one
\begin{equation} 
\label{eq:Fermi_energy_correction}
\hat H' = \hat H_0 -\lambda_{p}(\hat Z-Z_{0})- \lambda_{n}(\hat N-N_{0}),
\end{equation} 
where the Lagrange multiplier $\lambda_{\tau}$ is chosen to be the Fermi energy, which is determined by requiring the correct average particle number $\bra{\Phi(\mathbf{q})}\hat N_\tau\ket{\Phi(\mathbf{q})}=N_0(Z_0)$. It has been shown in Ref.~\cite{Yao:2011} that the deviation of the average particle number from the correct one can be as large as 0.4 particles, both for neutrons and protons. The subsidiary term could bring an evident correction to nuclear total energy. Moreover, this deviation displays a pronounced dependence on both the angular momentum and deformation.  One can see from Fig.~\ref{fig:Mg24_PNP}(a) that the violation of particle numbers in oblate states produces a wrong ordering of the projected energy curves for which the $J=0$ (red) one is even higher than the $J=6$ (pink). With the correction  considered in (\ref{eq:Fermi_energy_correction}), the energy ordering of the energy curves becomes normal and closer to the results of  the exact particle-number projection calculation.  
 
 Finally, PGCM calculation was carried out  by mixing all the axially deformed states, c.f. (\ref{eq:GCM_wf}), where the collective coordinate $\mathbf{q}$ is discretized.   One obtains discrete states with their energies indicated by dots centered at their mean deformations  $\bar{\beta}_{J, \sigma}$  defined as
 \begin{equation}
 \label{eq:average_beta_GCM}
 \bar{\beta}_{J,\sigma}=\sum_{\mathbf{q}_i} \beta(\mathbf{q}_i)\left|g_{J,\sigma}(\mathbf{q}_i)\right|^{2},
 \end{equation}
 where $g_{J,\sigma}$ has been defined in (\ref{eq:wf_g}).  It is shown in Fig.~\ref{fig:Mg24_PNP} that a rather well-defined rotational band ($0^+_1, 2^+_1, 4^+_1, 6^+_1$) is built upon the prolate deformed configurations around $\beta_{20}=0.5$. The energy spectrum of this rotational band is slightly more stretched in the calculation with the particle-number projection.

%\section{Applications to selected topics of nuclear structure}
 
  \subsection{The densities of symmetry-restored states}

\begin{figure}[tb]
\centering
\includegraphics[width=0.8\textwidth]{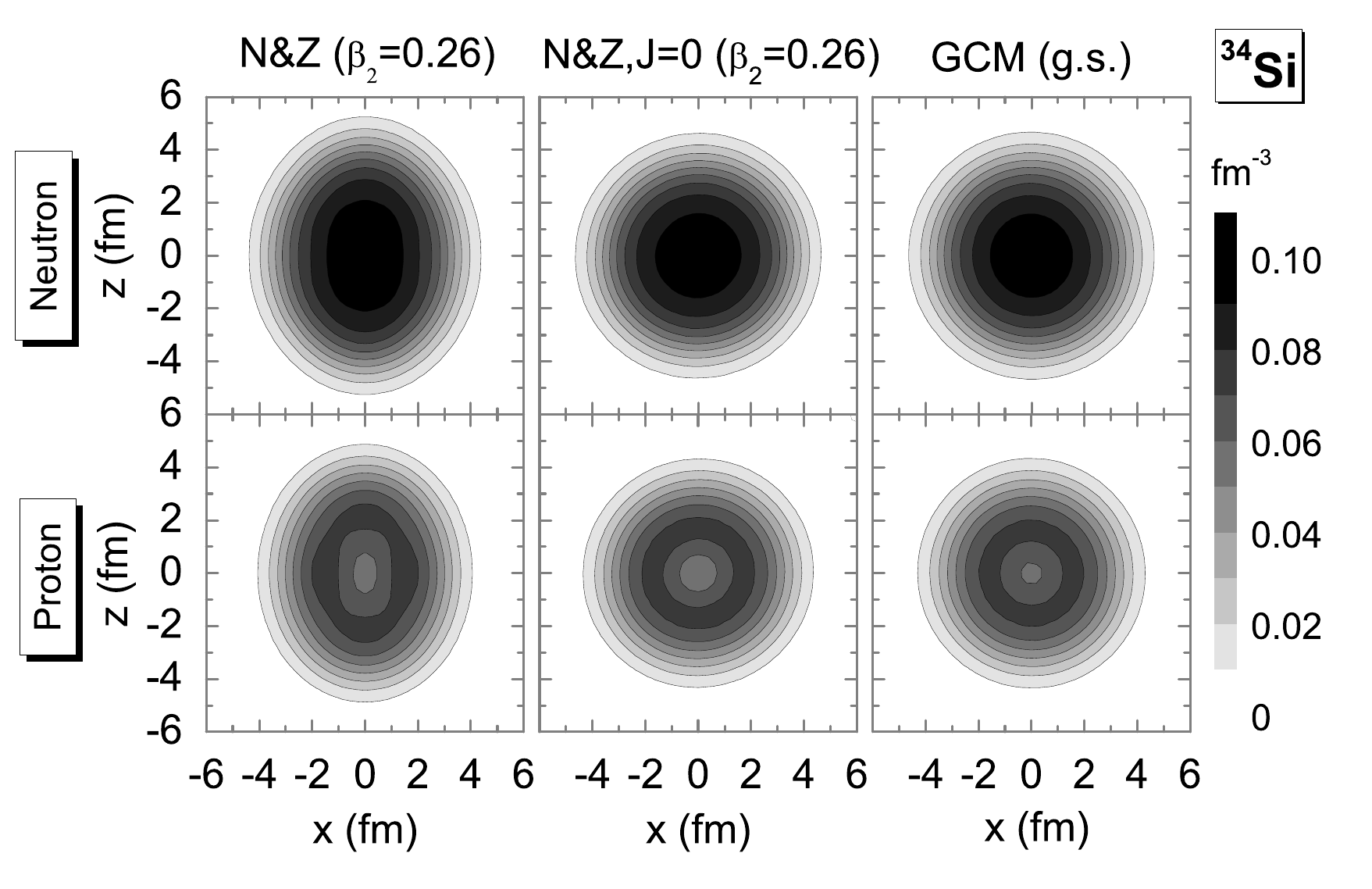}
\caption{ The contour plots of the neutron (upper panels) and proton (lower panels) densities of $^{34}$Si in the $y=0$ plane for the particle number projected HFB+Lipkin-Nogami state with $\beta_2=0.26$ (left column), its projection on both particle numbers and total angular momentum $J = 0$ (middle column) and for the $0^+_1$ GCM ground state (right column). The  Skyrme energy density functional SLy4~\cite{Chabanat:1997NPA} was employed in the calculations. Taken from Ref. ~\cite{Yao:2012PRC_Si34}. Figure reprinted with permission from the American Physical Society.
}
\label{fig:Si34_Density_gs}
\end{figure}

The effect of restoring angular momenta for nuclear low-lying states can be clearly visualized from their density distribution, 
\begin{eqnarray} 
\rho^{J}(\mathbf{r})
\equiv\bra{\Phi_{JN_0Z_0}(\mathbf{q})} \sum_{i} \delta\left(\mathbf{r}-\mathbf{r}_{i}\right) \ket{\Phi_{JN_0Z_0}(\mathbf{q})},
\end{eqnarray}
where index  $i$  runs over occupied single-particle states for neutrons or protons. The projected density $\rho^{J}(\mathbf{r})$ has the following form~\cite{Yao:2015PRC_Density}
\begin{eqnarray}
\label{eq:density_spin}
\rho^{J}(\mathbf{r})
&=& \sum_{L} \langle J 0 L 0 \mid J 0\rangle \rho^{J}_L(r)Y_{L 0}(\hat{\mathbf{r}}),
\end{eqnarray}
where $\hat{\mathbf{r}}$ stands for the angular part of the coordinate $\mathbf{r}$. The radial part $\rho^{J}_L(r)$ of the $L$-component of the density is defined as
\begin{eqnarray}
\label{eq:density_spin_J}
\rho^{J}_L(r)
=\sum_{K}(-1)^{K}\left\langle J K L-K \mid J 0\right\rangle 
\int d \hat{\mathbf{r}}  \rho_{J K}(r,\hat{\mathbf{r}}) Y_{L K}^{*}(\hat{\mathbf{r}})
\end{eqnarray}
with the quantity $\rho_{JK}(r,\hat{\mathbf{r}})$ given by
\begin{eqnarray}
\rho_{J K}(\mathbf{r})
&\equiv & \frac{2 J+1}{2} \int_{0}^{\pi} \sin\theta d\theta d_{K0}^{J *}(\theta) 
\bra{\Phi(\mathbf{q})}\sum_{i} \delta\left(\mathbf{r}-\mathbf{r}_{i}\right) e^{-i \theta \hat{J}_{y}} \hat{P}^{N_0} \hat{P}^{Z_0} \ket{\Phi(\mathbf{q})}.
\end{eqnarray}
The Eq.(\ref{eq:density_spin}) corresponds to the multipole expansion of the density  $\rho^J(\mathbf{r})$ of the projected state labeled with a definite angular momentum $J$.  As a special case, the density of the ground state with   $J=0$ is simplified as
\begin{eqnarray}
\rho^{J=0}(\mathbf{r}) 
&=&  \frac{1}{\sqrt{4\pi}} Y_{00}(\hat{\mathbf{r}})
\int d \hat{\mathbf{r}}^\prime   \rho_{00}(r,\hat{\mathbf{r}}^\prime ).
\end{eqnarray}

 Figure~\ref{fig:Si34_Density_gs} illustrates how the density distribution of neutrons (upper panels) and protons (lower panels) is modified at different levels of  calculation for $^{34}$Si. The left column shows contour plots of both neutron and proton densities for the particle-number projected HFB+Lipkin-Nogami state $\ket{\Phi(\mathbf{q})}$ with  $\beta_{2}=0.26$. Because of nonzero quadrupole deformation, the rotation symmetry is violated in the densities and a deformed semi-bubble structure is observed.  After projection on total angular momentum $J = 0$ (middle column), the density is obtained in the laboratory frame and becomes spherical. The configuration mixing in the PGCM calculation increases the central proton density again and simultaneously reduces the value at the bulge. As a result,  the depletion factor defined as
$F_{\max } \equiv (\rho_{\max, p}-\rho_{\mathrm{cent}, p})/\rho_{\max , p}$, which measures the reduction of the density at the nucleus center relatively to its maximum value, is reduced by the beyond-mean-field effect.

\subsection{Dynamical correlation energies}

Symmetry restoration is equivalent to the diagonalization of  nuclear Hamiltonian in a set of degenerate rotated states $\ket{\Phi(\mathbf{q}, \varphi)}$, which generally leads to an additional (dynamical) correlation energy correction to nuclear ground state. The correction energy from the restoration of rotation symmetry for the deformed state $\ket{\Phi(\mathbf{q})}$ can be calculated as follows,
\begin{equation}
    \Delta{\cal E}_{J=0}(\mathbf{q})) 
    = E(\mathbf{q}) - E_{J=0}(\mathbf{q})
    =\frac{\bra{\Phi(\mathbf{q})} \hat H \ket{\Phi(\mathbf{q})}}{\bra{\Phi(\mathbf{q})} \hat{\mathbb{1}} \ket{\Phi(\mathbf{q})}}
    -\frac{\bra{\Phi_{J}(\mathbf{q})} \hat H \ket{\Phi_{J}(\mathbf{q})}}{\bra{\Phi_{J}(\mathbf{q})} \hat{\mathbb{1}} \ket{\Phi_{J}(\mathbf{q})}}.
\end{equation}
Since this energy correction varies with nuclear deformation, the location of the energy minimum on the energy surface might be shifted after symmetry restoration. Given this situation, it is more natural to define the {\em rotational energy correction} as the difference between the energies of the lowest mean-field state $\ket{\Phi(\mathbf{q}_{\rm mf})}$ and the lowest angular-momentum projected state  $\ket{\Phi(\mathbf{q}_0)}$, 
\begin{equation}
    \Delta E_{J=0} = E(\mathbf{q}_{\rm mf}) - E_{J=0}(\mathbf{q}_0).
\end{equation}

\begin{figure}[tb] 
\includegraphics[width=0.5\textwidth]{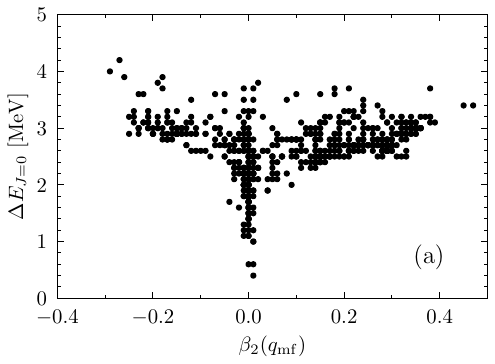}
\includegraphics[width=0.5\textwidth]{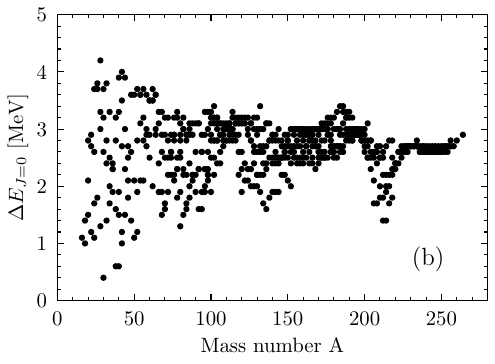}
\caption{Dynamical correction energy $\Delta E_{J=0}$ as a function of the (a) quadrupole deformation of the mean-field energy minimum state or (b)  the mass number $A$ for the 605 nuclei, where the mean-field states are generated from the HF+BCS calculation using the Skyrme interaction SLy4~\cite{Chabanat:1997NPA}. Results are taken from Ref.~\cite{Bender:2006Global}.
}
\label{fig:correlation_energy}
\end{figure}

The rotational energy correction $\Delta E_{J=0}$  for 605 nuclei from a MR-EDF calculation based on the SLy4 force is shown in Fig~\ref{fig:correlation_energy} as a function of the quadrupole deformation $\beta_2$ at the mean-field energy minimum or as a function of the nuclear mass number $A$, where both particle-number and angular momentum projection are implemented for axially deformed states. For simplification, the so-called topological Gaussian overlap approximation is employed for angular momentum projection~\cite{Bender:2006Global}.
%
% \begin{subequations}
%  \begin{align}
% \bra{\Phi(\mathbf{q})}\hat{R}_y(\theta)\ket{\Phi(\mathbf{q }}
% \simeq&\bra{\Phi(\mathbf{q})}\Phi(\mathbf{q})\rangle e^{-c_{2}\left(\mathbf{q}, \mathbf{q} \right) \sin^{2}\theta}, \\
% \bra{\Phi(\mathbf{q})}\hat H\hat{R}_y(\theta)\ket{\Phi(\mathbf{q }}
% \simeq&\bra{\Phi(\mathbf{q})}\Phi(\mathbf{q })\rangle e^{-c_{2}\left(\mathbf{q}, \mathbf{q} \right) \sin^{2}\theta}\nonumber \\
% & \times\left[h_{0}\left(\mathbf{q}, \mathbf{q} \right)-h_{2}\left( \mathbf{q}, \mathbf{q} \right) \sin^{2}\theta\right].
% \end{align}
% \end{subequations}
% The widths $c_{2}\left(\mathbf{q}, \mathbf{q}\right)$ of the Gaussian and the expansion coefficient $h_i\left(\mathbf{q}, \mathbf{q}\right)$ in the Hamiltonian kernel are determined from the matrix element in which the left state is rotated by a certain angle  $\theta$.   
It is seen that the $\Delta E_{J=0}$ is small for spherical/weakly deformed nuclei, and is about 3 MeV for either oblate or prolate deformed nuclei. Besides, the $\Delta E_{J=0}$s for  light nuclei spread in between 1 MeV and 4 MeV. In contrast, the $\Delta E_{J=0}$ is close to 3.0 MeV in heavy nuclei.  The size of energy correlation from symmetry restoration seems to be not sensitive to the employed energy density functionals~\cite{Rodriguez:2014mass,DRHBcMassTable:2022}.

% Later on, Rodriguez {\em et al.}~\cite{Rodriguez:2014mass} carried out a global study based on the Gogny D1S and D1M forces without the use of the topGOA. It was found that the inclusion of correction energies from both symmetry restoration and shape fluctuation can either improve (D1M) or worsen (D1S) the description of nuclear binding energies, depending on how the parameters of the forces are parameterized.  A recent study based on the CDFT~\cite{DRHBcMassTable:2022} has shown clearly that the inclusion of the rotational correction energy is essential to achieve a better description of nuclear masses with the PC-PK1 force, even though the rotational correction energy was estimated approximately  with the cranking approximation.

\subsection{Triaxiality in atomic nuclei with shape coexistence}

The PGCM provides a powerful tool of choice to study nuclei with strong shape mixing (the wave function of one nuclear state is an admixture of prolate and oblate deformed shapes) or  {\em shape coexistence}. The latter is a nuclear phenomenon that the low-lying states of an atomic nucleus consist of two or more states with similar energies which have well-defined and distinct properties and can be interpreted in terms of different intrinsic shapes.  The occurrence of shape coexistence affects the evolution behavior of nuclear low-lying states with  spin and isospin.  Neutron-deficient krypton isotopes are typical examples.

\begin{figure}[tb]
\centering
\includegraphics[width=\textwidth]{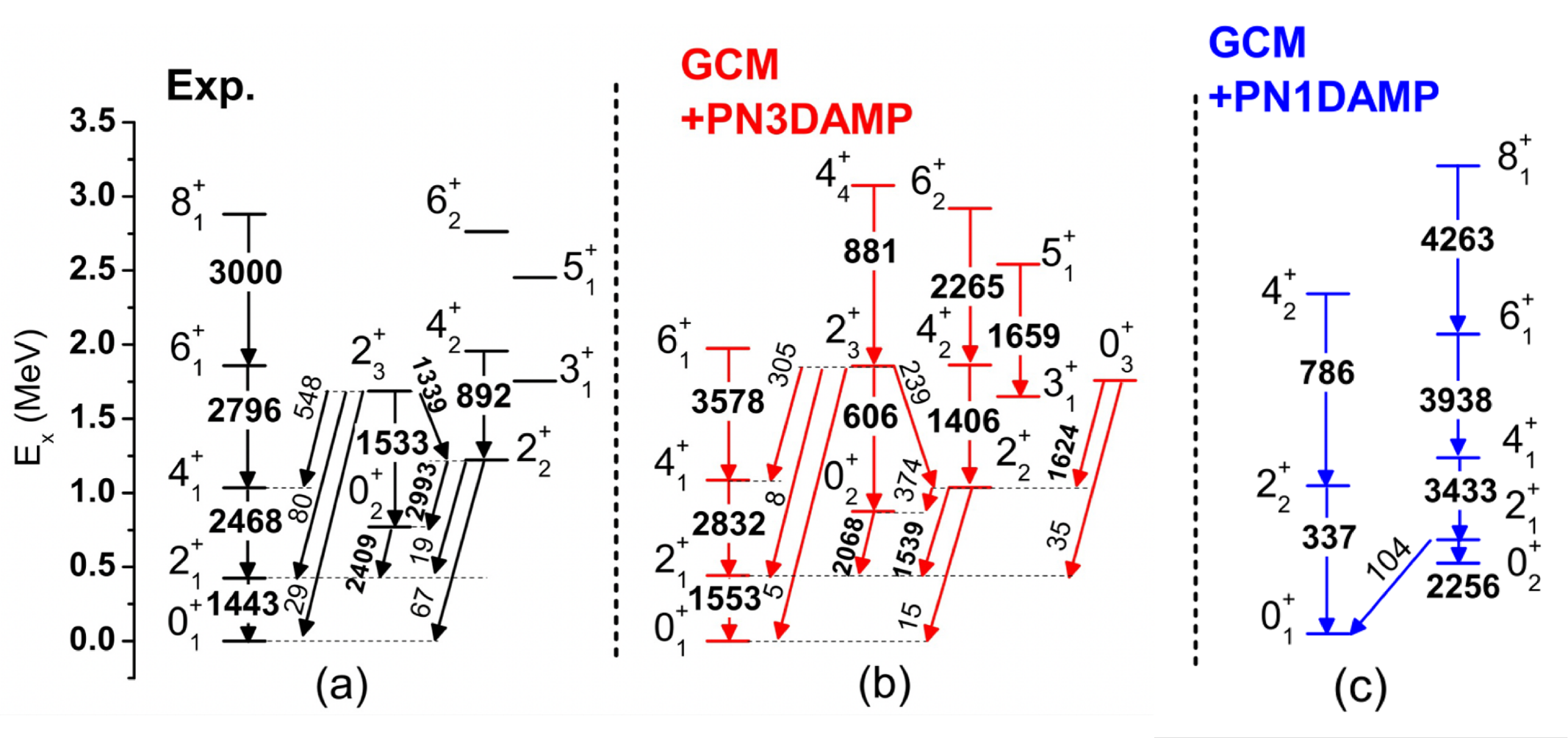}
\caption{Low-lying energy spectra of $^{76}$Kr from the MR-CDFT calculation with the mixing of (c) axially  and (b) triaxially ($\beta_{22}\neq 0$)  deformed states projected onto both particle numbers and different angular momenta, in comparison with data (a). The  electric quadrupole transition strengths $B(E2)$ (in $e^2$fm$^4$)  are  indicates on arrows. Adapted from Ref.~\cite{Yao:2014PRC}.
Figure reprinted with permission from the American
Physical Society.}
\label{fig:Kr76_spectra}
\end{figure}
 
Figure~\ref{fig:Kr76_spectra} displays the low-lying spectra of $^{76}$Kr from the PGCM calculation by mixing axially deformed only or also triaxially deformed configurations, in comparison with data.  It is shown that a restriction to axial states fails to reproduce the low-energy structures of the spectrum (including both excitation energies and electric quadrupole transitions), demonstrating the important role of including the triaxially deformed configurations in the description of the structure of the low-lying states of $^{76}$Kr.

\subsection{Dynamical and static octupole deformation}

The atomic nucleus $^{224}$Ra is usually suggested to be a stable pear-shaped nucleus based on the measured electric multiple ($E\lambda$) transitions~\cite{Ahmad:1993,Butler:1996}. According to the rotation energy spectrum of asymmetric diatomic molecular, one is expected to observe a rotational band with alternating parity for the states with even and odd angular momenta in $^{224}$Ra. However, this feature is shown only in the high-spin states of $^{224}$Ra, not in the low-spin states.  It has been interpreted as the rotation excitations of a quadrupole deformed ``rotor" on top of octupole vibration motions.

\begin{figure}[]
\centering
\includegraphics[width=\textwidth]{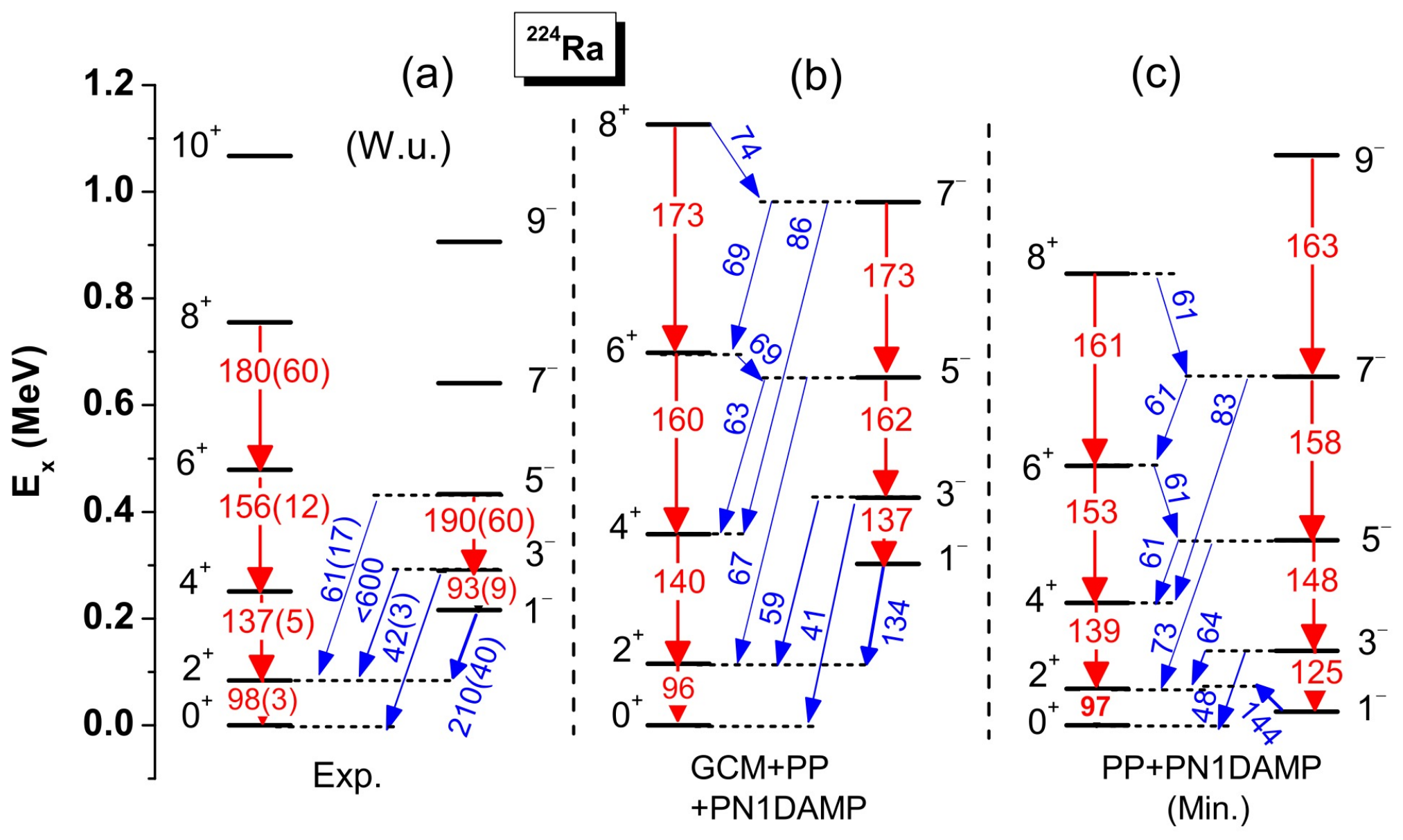}
\caption{Low-lying energy spectra for $^{224}$Ra. The results from full configuration and the single energy-minimum configuration calculations are compared to available data. The numbers on arrows are $E2$ (red [light gray] color) and E3 (blue [dark gray] color) transition strengths (Weisskopf units). Taken from Ref.~\cite{Yao:2015Octupole}. Figure reprinted with permission from the American Physical Society.}
\label{fig:Ra224_spectrum}
\end{figure}

  Figure~\ref{fig:Ra224_spectrum} shows the calculated low-lying energy spectra of  $^{224}$Ra. It is seen that the spectra and the $E2$, $E3$ transitions can be reproduced reasonably well using only the energy-minimum configuration. However, the energy displacement between the parity doublets in the low-spin region can only be reproduced in the PGCM calculation by mixing the configurations around the equilibrium shape. As shown in Ref.~\cite{Yao:2015Octupole}, the collective wave function of the ground-state is broadly distributed in the deformation plane, indicating a large shape fluctuation. With the increase of angular momentum, the collective wave functions of positive-parity states become gradually concentrated around the energy minimum and close to that of negative-parity states, displaying the classical picture of the stabilization of nuclear shapes with the increase of rotation frequency.

\subsection{Evolution of shell structure in neutron-rich nuclei}
  
The information on the evolution of shell structure can be learned from the properties of nuclear low-lying states. Generally speaking, a nucleus with a large shell gap is dominated by spherical  or weakly deformed shapes with high excitation energy of the first  $2^+$ state and low $E2$ transition strength between this state and the ground state. The description of the weakening of the neutron shell gap at $N=20$ is challenging for many nuclear models.  For example, the spectroscopic data for the low-lying states of $^{32}$Mg indicate that this nucleus is deformed in the ground state. However,   it is predicted to be spherical in almost all the mean-field approaches. 

%Previous studies have shown that the inclusion of beyond-mean-field effects from symmetry restoration and configuration mixing is essential to reproduce this peculiar phenomenon in this and other nuclei of the region. 
     
\begin{figure}[tb]
\centering
\includegraphics[width=\textwidth]{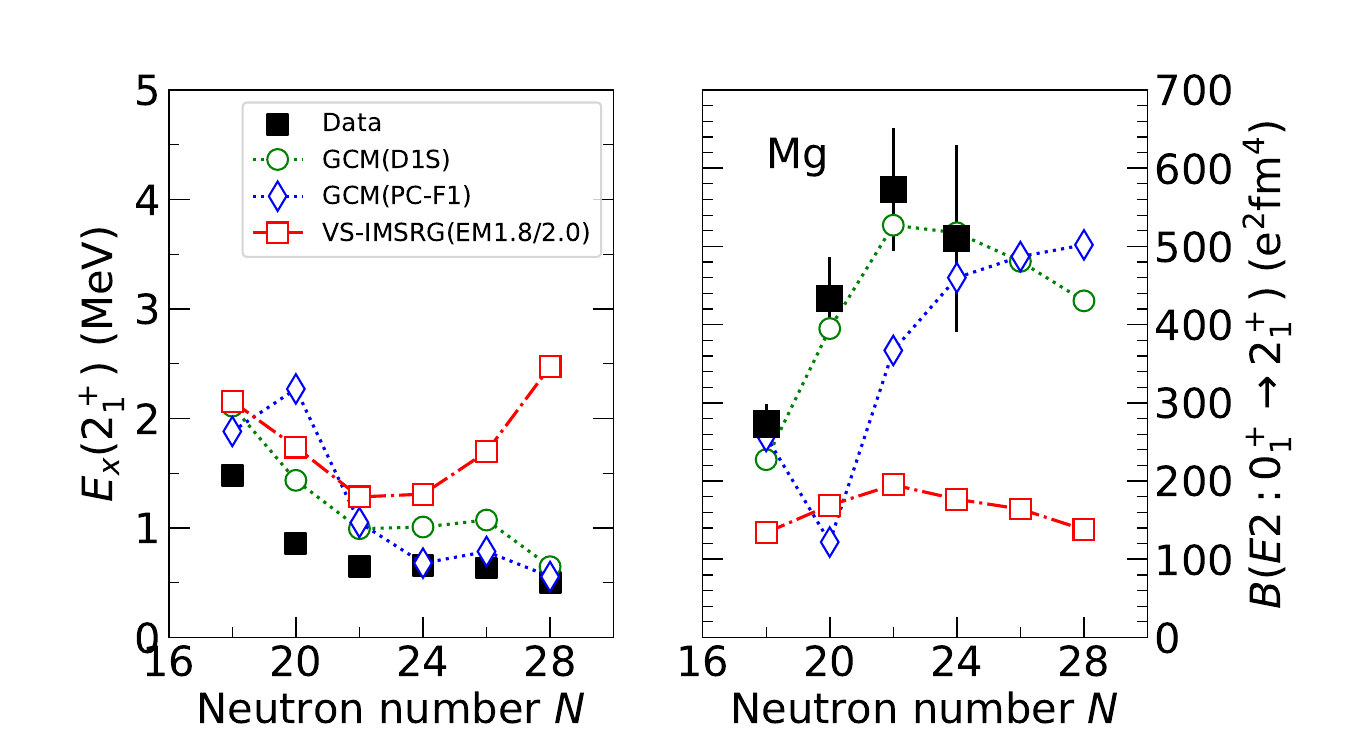}
\caption{Systematics of the excitation energy of $2^+_1$ state and the electric quadrupole transition strength of $0^+_1 \to 2^+_1$ in the neutron-rich magnesium isotopes from the PGCM calculations based on two different energy density functionals~\cite{Rodin:2006,Yao:2011} and the {\em ab initio} VS-IMSRG calculation based on the NO2B approximation~\cite{Miyagi:2020}, in comparison with available data~\cite{NNDC}.
}
\label{fig:Mg_isotopes}
\end{figure}

Figure~\ref{fig:Mg_isotopes} shows the systematic evolution of the excitation energy of $2^+_1$ state and the electric quadrupole transition strength of $0^+_1 \to 2^+_1$ in the neutron-rich magnesium isotopes from two multi-reference EDF calculations~\cite{Rodin:2006,Yao:2011} and the {\em ab initio} VS-IMSRG calculation~\cite{Miyagi:2020}.  The data show that the $N=20$ and $N=28$ shell gaps are melted in magnesium isotopes. This phenomenon has been well described by the non-relativistic EDF D1S force, partially by the relativistic EDF PC-F1 force, but poorly by the current implementation of valence-space IMSRG  based on the normal-ordering two-body approximation with a chiral NN+3N interaction. The description of nuclear large deformation and shape coexistence is still a challenge to nuclear {\em ab initio} methods.

\section{Concluding Remarks}

 Symmetry and group theory play an important role in modeling atomic nuclei. In many mean-field-based nuclear models, certain symmetries of nuclear Hamiltonian are allowed to be broken in mean-field potentials to incorporate some essential many-body correlations, which leads to a symmetry-breaking nuclear wave function. This symmetry breaking is introduced artificially because of the employed approximations and the broken symmetries need to be restored. In this chapter, the basic idea of this symmetry-breaking mechanism in mean-field approaches and the methods used to restore the broken symmetries at the beyond-mean-field level have been introduced with some illustrative examples. In the meantime, the success of these methods in the description of nuclear low-lying states has been demonstrated. 
 
%  It should be pointed out that symmetry-restoration methods have also been applied to study hypernuclear spectroscopy~\cite{Isaka:2011,Mei:2016,Cui:2017}, clustering structure in light nuclei~\cite{Zhou:2019}, and many other interesting topics which are not covered in this Chapter because of the page limit.

  Symmetry restoration has been popularly implemented in energy density functional calculations. However, most of the currently employed nuclear energy functionals were parameterized at the mean-field level. The inclusion of beyond-mean-field effects from symmetry restoration and also configuration mixing may worsen the description of nuclear properties. A new parametrization of energy density functionals including in the fitting procedure the beyond-mean-field effects is required to improve further the accuracy of the calculation with nuclear multi-reference  energy density functionals. Besides, it is worth mentioning that symmetry-restoration methods may meet the singularity and self-interaction problems if one starts from a general energy density functional~\cite{Anguiano:2001NPA,Bender:2009,Duguet:2009}. This discussion is not covered in this Chapter. Lots of efforts are devoted to finding out an energy functional free of these problems. This problem does not exist in the Hamiltonian-based approaches. Therefore, the implementation of symmetry-restoration methods into {\em ab initio}  methods starting from a realistic nuclear force derived from for instance the chiral effective field theory proposed by Weinberg~\cite{Weinberg:1991} becomes very attractive. The above-mentioned problems are not shown in these {\em ab initio} frameworks. This development has stimulated great research interest in the application of symmetry-restoration methods to nuclear structure and decay properties related to fundamental interactions and symmetries, such as the clustering structure in light nuclei, isospin-symmetry breaking correction in the superallowed $\beta$-decay, Schiff moment in octupole deformed nuclei and the nuclear matrix elements of neutrinoless double-beta decay.
  
\section*{Acknowledgments}

I would like to thank J. Meng for constructive discussion during the preparation of this manuscript, and I. Ivanov for his careful reading of it. Besides, I thank all my collaborators in the development of symmetry-restoration methods starting from various nucleon-nucleon interactions or energy density functionals, including B. Bally, M. Bender, J. Engel, Y. Fu, K. Hagino, P.-H. Heenen, H. Hergert, C.F. Jiao, Z. P. Li, H. Mei,  P. Ring, T. R. Rodriguez, D. Vretenar, X. Y. Wu, E.F. Zhou, and many others. This work was supported in part by  the National Natural Science Foundation of China (Grant No. 12141501) and the Fundamental Research Funds for the Central Universities, Sun Yat-sen University.

%\bibliographystyle{apsrev} 
%\bibliographystyle{unsrt}
%\bibliography{ref}

\end{document}